\documentclass[10pt,authoryear]{elsarticle}
\usepackage{graphics, graphicx, amsmath, amssymb}
\usepackage{subfigure}
\usepackage[utf8]{inputenc}
\usepackage[usenames,dvipsnames]{color}

\expandafter\let\csname equation*\endcsname\relax
\expandafter\let\csname endequation*\endcsname\relax
\usepackage{amsmath}
\usepackage{dsfont}
\usepackage[T1]{fontenc}

\definecolor{Red}{rgb}{1,0,0}

\newcommand{\ket}[1]{\ensuremath{\vert #1 \rangle}}

{\catcode`\|=\active
 \gdef\Braket#1{\begingroup
\mathcode`\|32768\let|\BraVert\left<{#1}\right>\endgroup}}
\def\BraVert{\egroup\,\mid\,\bgroup}

\newcommand{\Tr}{\mathrm{Tr}}

\newcommand{\re}{\mathrm{Re\,}}

\newcommand{\id}{\mathds{1}}

\newcommand{\dd}{\mathrm{d}}
\newcommand{\pt}{\partial}

\begin{document}

\begin{frontmatter}
\title{Quantum Control}
\date{\today}

\author[jülich]{Matthias M. M\"uller}
\address[jülich]{Peter Grünberg Institute -- Quantum Control (PGI-8), Forschungszentrum Jülich GmbH, D-52425 Germany}
\author[jülich,bologna,köln]{Tommaso Calarco}
\address[bologna]{Dipartimento di Fisica e Astronomia, Università di Bologna, 40127 Bologna, Italy}
\address[köln]{Institute for Theoretical Physics, University of Cologne, D-50937 Germany}
\author[jülich,köln]{Felix Motzoi}

\begin{abstract}Quantum Control is the art of steering the fragile dynamics of a quantum system in a precise way toward a desired final state or transformation. It is a cornerstone of many quantum technology applications and at the same time allows us to shed light on the fundamental connection between the energy structure and the time dynamics of quantum systems at the heart of some of the most interesting phenomena of quantum mechanics. In this chapter on quantum control, we introduce the control problem, go through basic error sources and methods to suppress these errors and then discuss state-of-the-art techniques to solve the control problem, like adiabatic methods, quantum optimal control and methods of machine learning.
\end{abstract}


\end{frontmatter}

\tableofcontents

\section{Key points and objectives}
\begin{itemize}
    \item The interplay of quantum systems with (specially designed) classical fields allows us to steer the dynamics of a quantum system toward a desired goal
    \item We describe the connection between controlled quantum dynamics and building blocks for quantum technology applications
    \item A key challenge in quantum control is the imprecision of the control pulse and the qubit description. We discuss error sources and strategies to mitigate these errors
    \item We show basic approaches for analytic and approximate tools to engineer control pulses
    \item Powerful numerical tools for quantum control rely on optimization -- relating to the field of quantum optimal control -- and the use of machine learning for data-driven control solutions
\end{itemize}
\section{Introduction}
It is fascinating to observe how the quantum mechanical state of individual systems evolves over time at a microscopic level. Modern experiments make it possible to generate, manipulate, and read out fragile states involving multiple entangled quantum mechanical degrees of freedom. These mechanisms also form the basis of quantum technologies, ranging from quantum computers and simulators to applications in sensor technology and communication [\cite{Koch2022}]. However, to truly make the leap from the laboratory to quantum technology devices, all these manipulations of the degrees of freedom in quantum mechanics must be performed with the utmost precision—and that is the task of quantum control. Here, the physical fundamentals, the theoretical description, and the mathematical structures of the time-evolution equations have to come together with various paths to control solutions, from analytical methods to numerical optimization, and to data-driven machine learning. In this chapter, we aim to convey a solid understanding of the challenge we face, to examine various types of control problems—based on optimization objectives and sources of error—using examples, and then to present a basic toolkit of methods that can help us achieve our goals. Let us start by defining what the control problem is.

\subsection{Control Problem}
\begin{figure}
	\centering
	\includegraphics[width=0.5\textwidth]{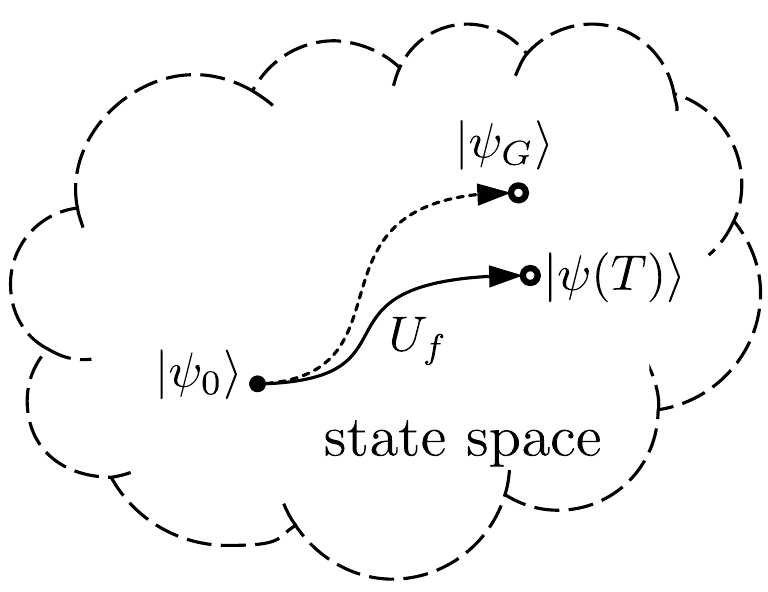}
	\caption{Control Problem. The system undergoes a time evolution (for example given by the unitary $U_f$ for closed systems) that depends on a control pulse $f(t)$, and steers the system, e.g., from an initial state $|\psi_0\rangle$ to a final state $|\psi(T)\rangle$ as illustrated by the solid arrow in the figure. The control problem is now to find $f(t)$ such that the final state of the system is as close as possible to the target state, here given by $|\psi_G\rangle$. The optimized trajectory is illustrated by the dashed arrow.}
\end{figure}
The core concept of quantum control is that a quantum system undergoes some dynamics that depends on the choice of an external control, e.g., a laser pulse, or a (time-modulated) electric or magnetic field, and this control is used to steer the system to a certain target, e.g., to reach a given target state or to realize a desired gate operation. Let us consider a quantum system that is described by a Hamiltonian 
\begin{eqnarray}
	H(t)=H_0(t)+H_c(f,t)\\
	\text{where, e.g. }H_c(f,t)=f(t)H_{c,0}\,,
\end{eqnarray}
where the time-dependent Hamiltonian $H(t)$ consists of a drift Hamiltonian $H_0(t)$ which is typically constant in time and a control Hamiltonian $H_c(f,t)$ that can depend on the time and on the control $f(t)$, often in a linear form $H_c(f,t)=f(t)H_{c,0}$. Setting $\hbar=1$, the time evolution of state $|\psi(t)\rangle$ is then given by the Schrödinger equation
\begin{eqnarray}
	|\dot\psi\rangle = -i H(t)|\psi(t)\rangle\,.
\end{eqnarray}
Formally, the evolution from initial time $0$ to final time $T$ of the state can then be encoded as $\ket{\psi(T)}=U(T)\ket{\psi(0)}$, where $U(T)$ is the appropriate unitary matrix. In more general terms, the quantum system could also be subject to various noise contributions and coupled to a larger system (or bath) . The time evolution would then be more complex to describe, e.g., by a Lindblad master equation, but the general concept of the control problem is the same [\cite{gorini1976,lindblad1976}].

The control problem is now to find a control such that the dynamics of the system is steered in the desired way. To make it more concrete, we consider two standard problems. First, we consider the \textbf{state transfer} problem. The task here is to start in a given initial state $|\psi_0\rangle=|\psi(0)\rangle$ and find the control pulse $f(t)$ such that at final time the system ends up in the target or goal state $|\psi_G\rangle$. To evaluate the performance of a control pulse $f(t)$ it is useful to quantify to which extent the final state $|\psi(T)\rangle$ corresponds to the target state $|\psi_G\rangle$. This can be done by any suitable state fidelity measure, e.g., the state overlap $F_{st}=|\langle \psi(T)|\psi_G\rangle|^2$, and the control problem is now to find a control pulse $f(t)$ such that $F_{st}$ is maximized (ideally $F_{st}=1$).
Another very important control problem is the generation of a desired \textbf{quantum gate} $V$. The time evolution according to the Schrödinger equation generates a gate $U(T)$ at final time that depends on the control pulse $f(t)$. This control pulse then has to be chosen such that $U(T)$ corresponds closely to $V$ in some metric. We will discuss different ways of obtaining such a correspondence later and limit ourselves here to the case where precisely $U(T)\hat{=}V$ up to a global phase. This can be quantified via the gate fidelity  
\begin{equation}
   F_{sm}=\frac{1}{N^2} \left|\sum_{n=1}^N \langle n | V^\dagger U(T)|n\rangle\right|^2=\frac{1}{N^2}|\Tr[ V^\dagger U(T)]|^2 .
\end{equation} Here, $N$ is the dimension of the Hilbert space and $\{|n\rangle\}_{n=1}^N$ are the computational basis states. The subscript "sm" stands for square modulus [\cite{Palao2008}]. Again, the control problem is to find a control pulse $f(t)$ such that $F_{sm}$ is maximized (ideally $F_{sm}=1$).

\section{Basic Control Techniques}
\subsection{Single-Qubit Gates and RWA}\label{sec:single-qubit}
A single qubit can be characterized by a level splitting or transition energy  $H_0=\frac{ \omega_0}2 \sigma_z$ and a control Hamiltonian $H_c(t)= \Omega(t) \cos(\omega_c t + \phi_c)\sigma_x$. Here, $\omega_0$ is the qubit frequency, $\sigma_{x,y,z}$ are the Pauli matrices, $\Omega(t)$ is the amplitude of the control pulse $f(t)=\Omega(t) \cos(\omega_c t + \phi_c)$, $\omega_c$ is the carrier frequency and $\phi_c$ the control phase. To analyze the dynamics of such a controlled two-level system, we go into the rotating frame by the transformation
\begin{eqnarray}
	|\tilde\psi(t) \rangle = e^{i\frac{ \omega_c}2 \sigma_z t }|\psi(t)\rangle\,.
\end{eqnarray}
This transformation of the state implies a transformation of the Hamiltonian given by
\begin{eqnarray}
	\tilde H(t) &=& e^{i\frac{\omega_c}2 \sigma_z t } H(t)e^{-i\frac{\omega_c}2 \sigma_z t } - \frac{\omega_c}2 \sigma_z \\
	&\approx&\frac{\delta}2 \sigma_z  + \frac{\Omega(t)}{2}\left(\cos\phi_c \sigma_x + \sin\phi_c \sigma_y\right)\,,
\end{eqnarray}
where $\delta=\omega_0-\omega_c$ is the detuning of the carrier frequency with respect to the qubit frequency. The approximation is called a Rotating-Wave Approximation (RWA) and is valid for $|\delta|,|\Omega(t)|\ll 2\omega_c$ and $\Omega(t)$ only slowly varying: $|\dot\Omega(t)/\Omega(t)|\ll \omega_c$ [\cite{rabi1937}].

For resonant driving ($\delta=0$) with an arbitrary control phase $\phi_c$ and rotation angle $\varphi_r=\int_0^t\Omega(t') dt'$, the unitary time evolution $U=e^{-i \tilde H t}$ takes the concise form
\begin{eqnarray}\label{eq:single-qubit-gate}
	U(\phi_c,\varphi_r)=\cos\frac{\varphi_r}{2}\id - i \left[\cos\phi_c\sigma_x + \sin\phi_c\sigma_y\right]\sin\frac{\varphi_r}{2}\,,
\end{eqnarray}
which is known as Rabi driving. This can be interpreted as a parameterized universal \textbf{single-qubit gate}. Indeed, we can construct a universal gate set by combining gates of the form Eq.~\eqref{eq:single-qubit-gate}. In particular, we have
\begin{itemize}
	\item $U(0,\pi)=-i\sigma_x = - i \begin{pmatrix}0&1\\1&0\end{pmatrix} \equiv X$
	\item $U(\pi/2,\pi)=-i\sigma_y = -i\begin{pmatrix}0&-i\\i&0\end{pmatrix} \equiv Y$
	\item $XY=-i\sigma_x (-i\sigma_y)=-i\sigma_z = -i\begin{pmatrix}1&0\\0&-1\end{pmatrix} \equiv Z$
	\item $U(\pi/2,\pi/2)Z=-i\frac{1}{\sqrt 2}\begin{pmatrix}1&1\\1&-1\end{pmatrix} \equiv H$, the Hadamard gate
	\item $H U(0,\pi/2)H=\frac{1}{2 \sqrt 2}\begin{pmatrix}1&1\\1&-1\end{pmatrix} \begin{pmatrix}1&-i\\-i&1\end{pmatrix}\begin{pmatrix}1&1\\1&-1\end{pmatrix}
	= \frac{1}{\sqrt 2}\begin{pmatrix}1-i&0\\0&1+i\end{pmatrix}= \begin{pmatrix}e^{-i\frac{\pi}{4}}&0\\0&e^{i\frac{\pi}{4}}\end{pmatrix}\equiv S$, the phase gate.
\end{itemize}

\subsection{Major Error Sources}
While it could seem that -- at least at the single-qubit level -- control problems are easy to solve, a more realistic description of quantum devices introduces additional challenges. The two-level system with perfectly-tuned parameters is an idealization or simplification and excludes all major error sources from the description. Starting from the description of the two-level Hamiltonian we can see that an uncertainty or instability in the  qubit frequency or carrier frequency will lead to an unknown detuning $\delta$ which can cause dephasing of the qubit. An uncertainty in the amplitude $\Omega(t)$ of the control pulse can lead to over- or under-rotation by altering the rotation angle $\varphi_r$. In larger systems with more than one qubit, the control pulse on one qubit can have a deleterious effect on other qubits if the individual addressing of the qubits is not perfect. This leads to an undesired dynamics of these other qubits, an error known as crosstalk. Additionally, errors can be caused by additional degrees of freedom in the system. Often, a qubit is formed by two levels out of a larger manifold of energy levels and leakage describes the fact that the population can leak out of the qubit subspace to these additional levels. In principle, there can be also external degrees of freedom that are not perfectly controllable and thus lead to an unavoidable average noise contribution.
While we describe all these effects for a single qubit, the concept is very similar for larger systems.

\subsection{Detuning and Robustness}\label{sec:detuning}
If the qubit frequency and the carrier frequency are not perfectly matched, e.g., due to uncertainties or fluctuations, the free evolution of the qubit is governed by
\begin{eqnarray}
	H = \frac{\delta}2 \sigma_z \,,
\end{eqnarray}
where the detuning $\delta$ is in principle unknown or even randomly fluctuating in time. To illustrate the effect of such a detuning on a qubit, we assume $\delta$ to be constant. If we prepare the qubit in a superposition state and let it evolve for a time $t$ we find
\begin{eqnarray}
\frac{|0\rangle + |1\rangle}{\sqrt 2} \stackrel{t}{\longrightarrow}
 \frac{e^{-i\frac{\delta}2  t}|0\rangle + e^{i\frac{\delta}2  t}|1\rangle}{\sqrt 2}
 =\cos\frac{\delta t}{2}|+\rangle - i \sin\frac{\delta t}{2}|-\rangle\,,
 \end{eqnarray}
 where we have introduced the superposition states $|+\rangle=(|0\rangle +|1\rangle) /\sqrt 2 $ and  $|-\rangle=(|0\rangle -|1\rangle) /\sqrt 2 $. The probability to find the system in its initial state $|+\rangle$ at time $t$ is then
\begin{eqnarray}
	p(|+\rangle)=\cos^2 \frac{\delta t}2 \,.
\end{eqnarray}
We can see that for small detunings or small times ($\delta t \ll 1$) the state is still close to the initial state, but if the time increases, the probability can go down to $0$. This becomes a problem especially if we do not know the precise value of $\delta$. To illustrate it, we assume that $\delta$ is randomly sampled from a Gaussian distribution around 0 with variance $\sigma_{\delta}^2$. The expectation value for the probability to find the system in its initial state is then calculated by averaging over $\delta$:
\begin{eqnarray}
	\langle p(|+\rangle) \rangle = \frac{1}{\sqrt{2\pi} \sigma_{\delta}} \int \cos^2 \frac{\delta t}2  \exp\left(-\frac{\delta^2}{2\sigma_{\delta}^2}\right) d \delta
				 = \frac 1 2\left( 1 + \exp\left(-\frac{\sigma_{\delta}^2 t^2}{2}\right) \right)\,.
\end{eqnarray}
We can see that for large times this probability converges to $1/2$ and we thus find the system randomly in $|+\rangle$ or $|-\rangle$ with the same probability. In other words, the phase information of the superposition between $|0\rangle$ and $|1\rangle$ is completely lost and we say that the system is dephased.

If we want to prevent the system from dephasing, we can apply control pulses that counter this dephasing. For this particular example, we can apply a so-called echo pulse in the middle of the time evolution. This so-called spin echo or Hahn echo [\cite{Hahn1950}] is ideally modeled via the instantaneous application of an $X$-gate halfway through the evolution (realistically, this can be realized to good approximation if $\Omega$ is much larger than $\delta$ so that the $X$-gate is fast compared to the dephasing induced by $\delta$). The time evolution is now given by
\begin{eqnarray}
|+\rangle \stackrel{t/2}{\longrightarrow}
\frac{e^{-i\frac{\delta}4  t}|0\rangle + e^{i\frac{\delta}4  t}|1\rangle}{\sqrt 2}
\stackrel{X}{\longrightarrow} \frac{e^{i\frac{\delta}4  t}|0\rangle + e^{-i\frac{\delta}4  t}|1\rangle}{\sqrt 2}  \stackrel{t/2}{\longrightarrow}  =|+\rangle\,.
\end{eqnarray}
Under the assumption of a constant detuning and a perfect and instantaneous echo gate we thus find that the initial state is perfectly preserved independently from the value of the detuning. This concept can be generalized to decouple different types of error by more complex pulse sequences and is known as dynamical decoupling~[\cite{Viola1999, Suter2016}].

\subsection{Amplitude Error}
The amplitude of a control pulse cannot always be tuned perfectly, e.g., due to miscalibration, thermal drift or inhomogeneities in a system with more than one qubit.
Starting from Eq.~\eqref{eq:single-qubit-gate}, we can see that an amplitude error $\Omega\rightarrow \Omega(1+\varepsilon)$ translates into an error in the rotation angle $\varphi_r\rightarrow \varphi_r(1+\varepsilon)$ and leads to a transformation
\begin{eqnarray}
U_\varepsilon(\phi_c,\varphi_r)=\cos\frac{\varphi_r(1+\varepsilon)}{2}\id - i \left[\cos\phi_c\sigma_x + \sin\phi_c\sigma_y\right]\sin\frac{\varphi_r(1+\varepsilon)}{2}\,.
\end{eqnarray}
For the $X$-gate ($\phi_c=0$, $\varphi_r=\pi$) this simplifies to
\begin{eqnarray}\label{eq:pulse_rotation_error}
	U_{\varepsilon}(0,\pi)=e^{-i\frac{\pi}{2}(1+\varepsilon) \sigma_x}\,.
\end{eqnarray}
If we want to use this gate, e.g., to perform a state transfer from $|0\rangle$ to $|1\rangle$, we find a fidelity
\begin{eqnarray}
	F_{st}=|\langle 1|U_{\varepsilon}(0,\pi)|0\rangle|^2=\sin^2\frac{\pi(1+\varepsilon)}{2}= \cos^2\frac{\pi\varepsilon}{2}\approx 1 - \frac{\pi^2\varepsilon^2}{4}\,
\end{eqnarray}
for a small rotation error $\varepsilon$.
To obtain some intuition into how to compensate for such an error in the control amplitude, we consider the identity
\begin{eqnarray}\label{eq:simple-composite}
    e^{-i\frac{\pi}{2} \sigma_x}=e^{-i\frac{\pi}{4} \sigma_y}e^{-i\frac{\pi}{2} \sigma_x}e^{-i\frac{\pi}{4} \sigma_y}\,.
\end{eqnarray}
We can interpret this as realizing the desired $X$-gate through a sequence of three control pulses around the $x$-axis ($\phi_c=0$) and $y$-axis ($\phi_c=\pi/2$). If all three of these pulses are subject to the same amplitude error, the transformation with error becomes
\begin{eqnarray}\label{eq:pulse_rotation_error_YXY}
		U_{\varepsilon}=e^{-i\frac{\pi}{4}(1+\varepsilon)  \sigma_y}e^{-i\frac{\pi}{2} (1+\varepsilon) \sigma_x}e^{-i\frac{\pi}{4} (1+\varepsilon) \sigma_y}\,.
\end{eqnarray}
By a simple calculation we can separate the desired transformation from the error contribution to obtain
\begin{eqnarray}\label{eq:pulse_rotation_error_YXY2}
	U_{\varepsilon}=e^{-i\frac{\pi}{2}\sigma_x} \cdot e^{i\frac{\pi}{4}\varepsilon  \sigma_y}e^{-i\frac{\pi}{2} \varepsilon \sigma_z}e^{-i\frac{\pi}{4} \varepsilon \sigma_y}\,.
\end{eqnarray}
With help of the Baker-Campbell-Hausdorff (BCH) formula we can simplify the error contribution for a small rotation error $\varepsilon$:
\begin{eqnarray}\label{eq:pulse_rotation_error_YXY3}
e^{i\frac{\pi}{2}\sigma_x} U_{\varepsilon}= e^{i\frac{\pi}{4}\varepsilon  \sigma_y}e^{-i\frac{\pi}{2} \varepsilon \sigma_z}e^{-i\frac{\pi}{4} \varepsilon \sigma_y}\approx e^{-i\frac{\pi}{2} \varepsilon \sigma_z}e^{i\frac{\pi^2}{4} \varepsilon^2 \sigma_x}\,.
\end{eqnarray}
As a result, this sequence of three control pulses (or composite pulse) gives us the same state preparation with an improved fidelity
\begin{eqnarray}
	F_{st}=|\langle 1|U_{\varepsilon}|0\rangle|^2\approx |\langle 0|  e^{-i\frac{\pi}{2} \varepsilon \sigma_z}e^{i\frac{\pi^2}{4} \varepsilon^2 \sigma_x} |0\rangle|^2 = \cos^2\frac{\pi^2\varepsilon^2}{4}\approx 1-\frac{\pi^4\varepsilon^4}{16} \,.
\end{eqnarray}
We have thus not only managed to suppress the error by a factor, but to turn a quadratic scaling in a quartic scaling of the state preparation error $1-F_{st}$ with the amplitude error $\varepsilon$.

\begin{figure}
    \centering
    \includegraphics[width=0.45\textwidth]{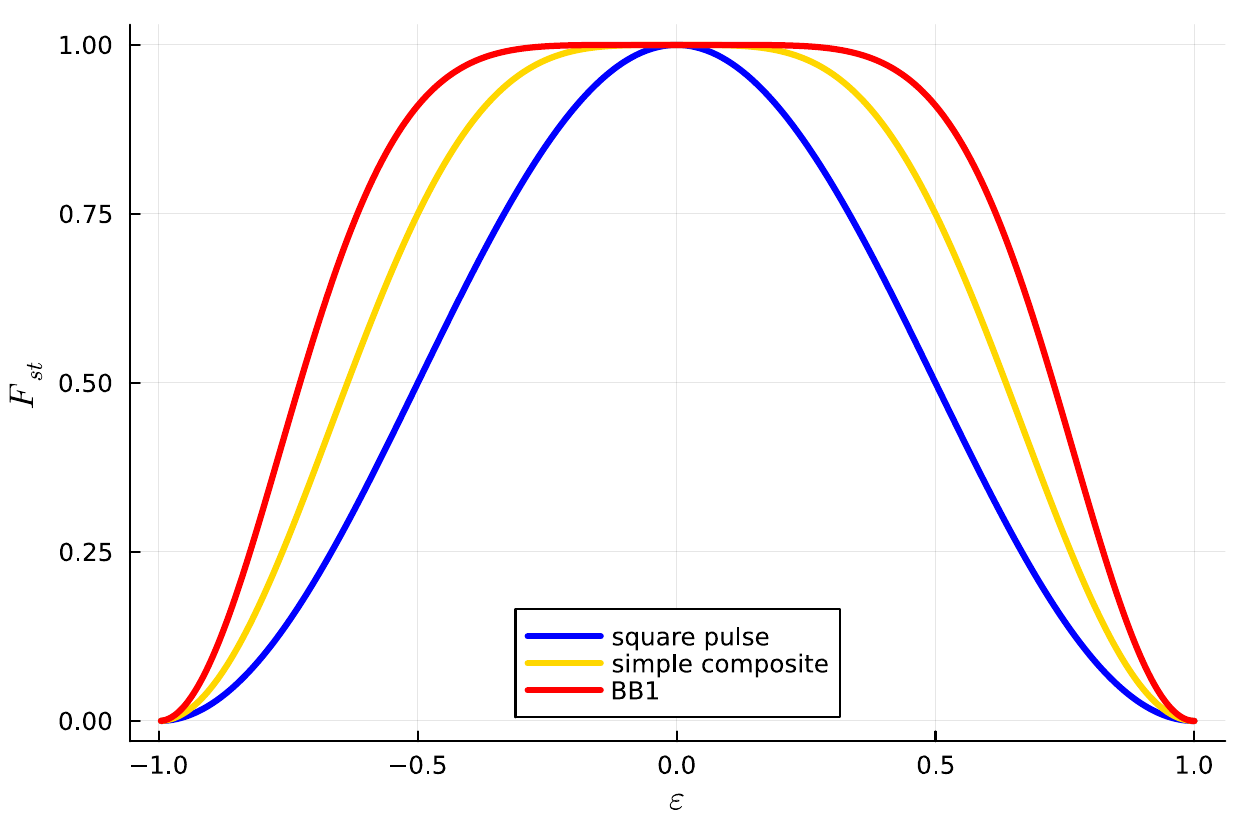}
    \includegraphics[width=0.45\textwidth]{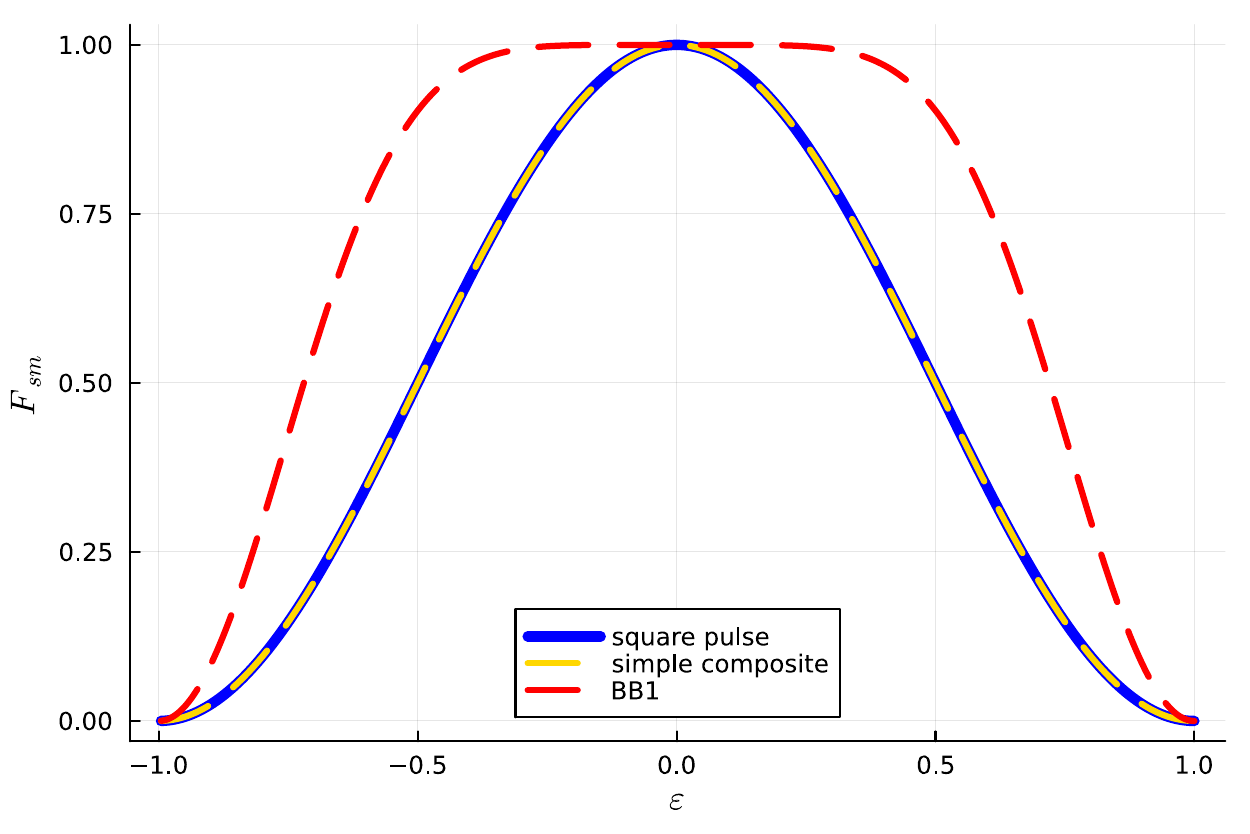}
    \caption{Compensation of amplitude errors for an $X$ gate. The simple composite pulse and BB1 lead to the same improvement in the state transfer fidelity (left panel) compared to the square pulse. For the gate fidelity, however, BB1 is still robust, while the simple composite pulse does not bring an advantage as it simply transforms the amplitude error into a phase error.}
    \label{fig:composite-pulses}
\end{figure}

What we can learn from this example is that an alternation of pulses along different axes can lead to a reduction of the overall error. This has been more systematically investigated, e.g., by \cite{Wimperis1994} who introduced several broad-band (BB) composite pulses such as BB1 which consists of three additional pulses along with the desired rotation:
\begin{eqnarray}\label{eq:BB1}
	U_{BB1,0}(\phi_c) e^{-i\frac{\varphi_r}{2} \sigma_x}= e^{-i\frac{\varphi_r}{2} \sigma_x}\,\\
	U_{BB1,0}(\phi_c)=U(\phi_c,\pi)	U(3\phi_c,2\pi)	U(\phi_c,\pi) = \id\,,
\end{eqnarray}
where the control phase has to be chosen as $\phi_c=\arccos(-\varphi_r/4\pi)$ to correct for the rotation error. Indeed, one can show that 
\begin{eqnarray}
	U_{BB1,\varepsilon}(\phi_c) e^{-i\varphi_r(1+\varepsilon) \sigma_x}\approx e^{-i\frac{\varphi_r}{2} \sigma_x}\,,
\end{eqnarray}
with an error of order $\varepsilon^3$ (leading thus to an error on the gate fidelity $F_{sm}$ of the order of $\varepsilon^6$). Fig.~\ref{fig:composite-pulses} shows the error dependence of the state transfer fidelity $F_{st}$ (left panel) and of the gate fidelity $F_{sm}$ (right panel) for a square pulse (constant $\Omega$), the simple composite pulse, Eq.~\eqref{eq:simple-composite}, and the BB1 pulse, Eq.~\eqref{eq:BB1}. As one can see, the simple composite pulse is more robust against the pulse error for state transfer compared to the square pulse, but for the gate fidelity this advantage is lost since the rotation error was simply transferred into a phase error, see Eq.~\eqref{eq:pulse_rotation_error_YXY3}. The BB1 pulse instead yields the enhanced robustness for both state preparation and gate fidelity.

Similar pulse sequences can be found also for the case where the error is not on the amplitude of the control pulse but on the frequency, e.g., the CORPSE (Compensation for Off-Resonance with a Pulse SEquence)~[\cite{Cummins2000,Cummins2003}] pulse sequence. The basic principle that commuting errors of different types (e.g., rotation errors around different axes) can compensate each other can be exploited also via numerical methods like the ones described in Sec.~\ref{sec:algorithms}.

\subsection{Crosstalk}

As an example for an error that arises  in systems that are larger than just one qubit, we discuss the problem of crosstalk. If we want to address a specific qubit, the control pulses might affect also other qubits. The individual addressability can be realized by focusing the control lines (i.e., laser beams) on single qubits -- crosstalk then arises when stray fields (weakly) drive an undesired qubit. Another approach that we discuss here is spectral addressability. In this case, all qubits have a different transition energy and the control pulse is on resonance only with one of the qubits (the one we want to drive). The other qubits are subject to off-resonant driving.

\subsubsection{Crosstalk for two strongly driven qubits}
As first example, we consider here two qubits with the Hamiltonians 
\begin{eqnarray}
	H_1=\frac{\Omega}{2}\sigma_x\\
	H_2=\frac{\Delta}{2}\sigma_z + \frac{\Omega}{2}\sigma_x\,
\end{eqnarray}
where the difference in the transition energies is 	$\Delta$ and the effective Rabi frequencies are $\Omega$ for the first qubit, and $\tilde \Omega_2=\sqrt{\Omega^2+\Delta^2}$ for the second qubit.

We want to implement $V_1=\sigma_x$ on the first qubit and $V_2=\id$ on the second qubit. We obtain the desired rotation angle on the first qubit by setting $\Omega t=\pi$:
\begin{eqnarray}
	U_1=-i\sigma_x=-i V_1\,.
\end{eqnarray}
The fidelity for implementing the desired gate $V_1$ is thus perfect:
\begin{eqnarray}
	F_{sm,1}=\frac{1}{4}|\Tr V_1^\dagger U_1|^2=1\,.
\end{eqnarray}
However, the second qubit is not untouched. It rather undergoes the time evolution
\begin{eqnarray}
	U_2=\cos\frac{\tilde\Omega_2 t}{2}\id - i \frac{\Omega\sigma_x +\Delta \sigma_z}{\tilde\Omega_2}\sin\frac{\tilde\Omega_2 t}{2}\,.
\end{eqnarray}
The fidelity for keeping the second qubit is
\begin{eqnarray}\label{eq:qubitrecurrence}
	F_{sm,2}=\frac{1}{4}|\Tr V_2^\dagger U_2|^2=\cos^2\frac{\tilde\Omega_2 t}{2}\,.
\end{eqnarray}
We can see that this fidelity oscillates between $0$ and $1$ and we have to choose $\tilde\Omega_2$ such that $F_{sm,2}=1$ while maintaining the choice $\Omega t=\pi$ that we needed for $F_{sm,1}=1$. A solution to this problem is $\Omega=\frac{\Delta}{\sqrt{4n^2 - 1}}$ and $t=\pi/\Omega$ for integer $n$.

More generally, the cross-talk problem can be perturbatively treated for an arbitrary number of qubits in the limit of weak driving and by considering the time-dependent nature of the field, as in the next problem. However, this weak-driving regime limits the gate speed at the expense of  further accumulation of decoherence effects.

\subsubsection{Time-dependent fields and AC-Stark / Bloch-Siegert shifts}\label{sec:blochsiegert}

So far, we have mainly considered resonant driving with a constant Rabi frequency. 
However, in real gate implementations the control amplitudes are time dependent,
$\Omega(t)$, and off-resonant couplings then induce additional phase shifts even when the undesired flip of the qubit is negligible. 
These shifts are commonly referred to as AC-Stark shifts (or Bloch-Siegert shifts, depending on the origin of the correction) [\cite{bloch1940,autler1955}]. 
As a consequence, even if the pulse suppresses unwanted transitions, the computational states may still acquire additional phases that must be compensated.

To illustrate the effect, we again consider the off-resonantly driven qubit,
\begin{eqnarray}
H
=
\frac{\Delta}{2}\sigma_z
+
\frac{\Omega(t)}{2}\sigma_x,
\end{eqnarray}
now under the assumption $|\Omega(t)|\ll \Delta$, i.e., that the drive is far off resonant and population transfer is strongly suppressed. 
Nevertheless, the drive still perturbs the energy levels and thereby induces phase shifts. This is precisely the regime relevant for weak crosstalk.

For a constant drive amplitude, the Hamiltonian can simply be diagonalized exactly, yielding the dressed energies
\begin{eqnarray}
E_\pm
=
\pm \frac{1}{2}\sqrt{\Delta^2+\Omega^2}.
\end{eqnarray}
Expanding for weak driving,
\begin{eqnarray}
\sqrt{\Delta^2+\Omega^2}
=
\Delta
\sqrt{1+\frac{\Omega^2}{\Delta^2}}
\approx
\Delta
+
\frac{\Omega^2}{2\Delta},
\end{eqnarray}
we see that the energy splitting is shifted by
\begin{eqnarray}
\delta_{\rm Stark}
=
\frac{\Omega^2}{2\Delta}.
\end{eqnarray}
Equivalently, the effective Hamiltonian becomes
\begin{eqnarray}
H_{\rm eff}
=
\left(
\frac{\Delta}{2}
+
\frac{\Omega^2}{4\Delta}
\right)\sigma_z.
\end{eqnarray}

Physically, this shift originates from virtual off-resonant excitations. 
In atomic physics, such shifts induced by an oscillating electric field are known as AC-Stark shifts. 
In driven two-level systems, related corrections arising from counter-rotating terms are often referred to as Bloch-Siegert shifts.

We will now more rigorously treat this regime of weak cross talk for time-dependent pulses. The condition
\begin{eqnarray}
|\Omega(t)|\ll \Delta
\end{eqnarray}
allows us to perturbatively eliminate the off-resonant coupling at each instant in time. 
 A convenient method is the Schrieffer--Wolff (SW) transformation [\cite{Schrieffer1966}], which approximately diagonalizes the Hamiltonian order by order in $\Omega/\Delta$.

We therefore perform the unitary transformation
\begin{eqnarray}
H_{\rm eff}
=
e^S H e^{-S},
\end{eqnarray}
with anti-Hermitian generator
\begin{eqnarray}
S(t)
=
i\frac{\Omega(t)}{2\Delta}\sigma_y.
\end{eqnarray}
Neglecting the derivative contribution $i\dot S$, which is of higher order in the adiabatic expansion and discussed in the next section, we use the BCH expansion
\begin{eqnarray}
H_{\rm eff}
=
H+[S,H]+\frac{1}{2}[S,[S,H]]+\cdots\\
=H_0+H_{\rm eff}^{(1)}+H_{\rm eff}^{(2)}+\cdots\,,
\end{eqnarray}
where the second row groups together different orders in the perturbation parameter.

Splitting the Hamiltonian into $H=H_0+V$ with
\begin{eqnarray}
H_0=\frac{\Delta}{2}\sigma_z,
\qquad
V=\frac{\Omega(t)}{2}\sigma_x,
\end{eqnarray}
the generator is chosen such that
\begin{eqnarray}\label{eq:SWcondition}
[S,H_0]
=
-V.
\end{eqnarray}

The leading correction then arises from the second-order terms in the BCH expansion,
\begin{eqnarray}
H_{\rm eff}^{(2)}
=
[S,V]
+
\frac{1}{2}[S,[S,H_0]].
\end{eqnarray}
Using Eq.~\eqref{eq:SWcondition}, the two contributions partially cancel and we obtain
\begin{eqnarray}
H_{\rm eff}^{(2)}
=
\frac{1}{2}[S,V].
\end{eqnarray}
Thus, we find
\begin{eqnarray}
[S,V]
=
i\frac{\Omega^2}{4\Delta}
[\sigma_y,\sigma_x]
=
\frac{\Omega^2}{2\Delta}\sigma_z,
\end{eqnarray}
and therefore
\begin{eqnarray}
H_{\rm eff}^{(2)}
=
\frac{\Omega^2(t)}{4\Delta}\sigma_z.
\end{eqnarray}
Thus, to second order in $\Omega/\Delta$, the effective Hamiltonian becomes
\begin{eqnarray}
H_{\rm eff}
=
\left(
\frac{\Delta}{2}
+
\frac{\Omega^2(t)}{4\Delta}
\right)\sigma_z.
\end{eqnarray}
We therefore obtain the time-dependent detuning
\begin{eqnarray}
\delta_{\rm Stark}(t)
=
\frac{\Omega^2(t)}{2\Delta}.
\end{eqnarray}
Continuing the BCH expansion can likewise produce higher order Stark shift terms.

For a constant pulse this reproduces the perturbative expansion of the exact eigenenergies above. 
For shaped pulses, however, the induced detuning becomes time dependent and leads to the accumulated phase
\begin{eqnarray}
\phi_{\rm Stark}(t)
=
\int_0^t dt'\,
\delta_{\rm Stark}(t').
\end{eqnarray}

Consequently, even weak off-resonant driving can induce coherent $Z$-errors during gate operations. 
These phase shifts must therefore be compensated  [\cite{motzoi2013}], for example by adding a suitable counter-detuning or by dynamically adjusting the drive phase, for example via subsequent (virtual) Z rotations.

\subsection{Leakage}\label{sec:leakage}

Finally, we now consider the effect of leakage to higher excited states. For simplicity, we restrict ourselves to the three lowest levels ${|0\rangle,|1\rangle,|2\rangle}$ and assume a weakly anharmonic system (i.e.~couplings only between adjacent states). In the rotating frame of the drive (with respect to the $|0\rangle \leftrightarrow |1\rangle$ transition), the Hamiltonian can be written as
\begin{eqnarray}
H(t) = - \Delta |2\rangle\langle 2|+
\frac{1}{2} \left[
\Omega_I(t) I + \Omega_Q(t) Q
\right],
\end{eqnarray}
where $\Delta$ denotes the anharmonicity and we have introduced the operators
\begin{eqnarray}
 I&=
\begin{pmatrix}
0 & 1 & 0 \\
1 & 0 & \lambda_{12} \\
0 & \lambda_{12} & 0
\end{pmatrix}, \qquad
Q =
\begin{pmatrix}
0 & -i & 0 \\
i & 0 & -i\lambda_{12} \\
0 & i\lambda_{12} & 0
\end{pmatrix}.
\end{eqnarray}
The terms proportional to $\lambda_{12}$ describe the unwanted coupling between $|1\rangle$ and $|2\rangle$, which take the value $\sqrt{2}$ in the harmonic oscillator limit or $1$ for a spin-1 system.

To analyze leakage, we move to the interaction picture with respect to
\begin{eqnarray}
H_0 = -\Delta |2\rangle\langle 2|.
\end{eqnarray}
The coupling operators then acquire a time dependence
\begin{eqnarray}
|1\rangle\langle 2| \rightarrow |1\rangle\langle 2| e^{i\Delta t}, \qquad
|2\rangle\langle 1| \rightarrow |2\rangle\langle 1| e^{-i\Delta t},
\end{eqnarray}
so that the interaction Hamiltonian becomes
\begin{eqnarray}
H_I(t) = \frac{1}{2} \Big[
\Omega_I(t)\big( I_{01} + \lambda_{12} I_{12}(t) \big)
+
\Omega_Q(t)\big( Q_{01} + \lambda_{12} Q_{12}(t) \big)
\Big],
\end{eqnarray}
where 
\begin{eqnarray}
I_{01}&=&|0\rangle\langle1|+|1\rangle\langle0|,\\
Q_{01}&=&-i|0\rangle\langle1|+i|1\rangle\langle0|.
\end{eqnarray}
and $I_{12}(t),Q_{12}(t)$ denote the oscillating $|1\rangle \leftrightarrow |2\rangle$ terms.

\subsubsection{First-order Magnus (average Hamiltonian)}

To leading order, the time evolution is governed by the integral of the Hamiltonian, also known as the average Hamiltonian (or more formally the first term of the Magnus expansion [\cite{magnus1954}]),
\begin{eqnarray}
\bar{H}^{(1)} = \frac{1}{t} \int_0^t H_I(t') dt'.
\end{eqnarray}

The desired dynamics is contained in the $|0\rangle \leftrightarrow |1\rangle$ subspace. To suppress leakage, we require that the contribution coupling to $|2\rangle$ vanishes in the average Hamiltonian. This leads to the condition
\begin{eqnarray}
\int_0^T dt \left[\Omega_I(t) - i \Omega_Q(t) \right] e^{i\Delta t} = 0.
\end{eqnarray}
This equation expresses that the spectral weight of the drive at frequency $\Delta$ must vanish, i.e., the pulse should not excite the off-resonant transition, and reduces to the same condition as Eq.~\eqref{eq:qubitrecurrence} for constant drive.

\subsubsection{Integration by parts and DRAG}

We now assume that the pulse envelope is smooth and vanishes at the boundaries,
\begin{eqnarray}
\Omega_I(0)=\Omega_I(T)=0\,.
\end{eqnarray}
We first consider the contribution from $\Omega_I(t)$:
\begin{eqnarray}
\int_0^T dt \Omega_I(t) e^{i\Delta t}.
\end{eqnarray}

Performing integration by parts yields
\begin{eqnarray}
\int_0^T dt \Omega_I(t) e^{i\Delta t}=
\left.\frac{\Omega_I(t)}{i\Delta} e^{i\Delta t}\right|_0^T
-\frac{1}{i\Delta} \int_0^T dt \dot{\Omega}_I(t) e^{i\Delta t}\,.
\end{eqnarray}
Due to the boundary conditions, the first term vanishes and we obtain
\begin{eqnarray}
\int_0^T dt\, \Omega_I(t) e^{i\Delta t}=-
\frac{1}{i\Delta} \int_0^T dt\, \dot{\Omega}_I(t) e^{i\Delta t}.
\end{eqnarray}
Inserting this into the leakage condition, we find
\begin{eqnarray}
\int_0^T dt\,
\left[_
-\frac{1}{i\Delta} \dot{\Omega}_I(t)
- i \Omega_Q(t)
\right] e^{i\Delta t}
=0.
\end{eqnarray}
This condition is satisfied if the integrand vanishes, which leads to
\begin{eqnarray}
\Omega_Q(t) = \frac{\dot{\Omega}_I(t)}{\Delta}\,.
\end{eqnarray}
This solution reproduces the Derivative Removal by Adiabatic Gate (DRAG) pulse sequence, which can likewise also be obtained by time-dependent adiabatic expansion of the dynamics, see \cite{Motzoi2009, motzoi2013}.

\subsubsection{Time-dependent Stark-shift correction}

The derivative quadrature derived above suppresses leakage transitions to first order. 
However, as discussed previously for the two-level system, off-resonant couplings also induce crucial AC-Stark shifts within the computational subspace. 
These shifts arise from higher-order terms in the Magnus expansion and therefore accompany the DRAG correction.

As before, the relevant contribution is generated by the second-order Magnus term
\begin{eqnarray}
-i t \bar H^{(2)}(t)
=
-\frac{1}{2}
\int_0^t dt_1
\int_0^{t_1}dt_2\,
[H_I(t_1),H_I(t_2)].
\end{eqnarray}
Equivalently, one may derive the same result using a Schrieffer--Wolff transformation as was done for crosstalk. 
To leading order in $1/\Delta$, both approaches give identical effective Hamiltonians. 
For simplicity, we neglect terms quadratic in the small DRAG quadrature $\Omega_Q$.

The first contribution arises from the computational subspace operators $I_{01}$ and $Q_{01}$.
Their commutator evaluates to
\begin{eqnarray}
[I_{01},Q_{01}]
=
2i
\left(
|0\rangle\langle0|
-
|1\rangle\langle1|
\right).
\end{eqnarray}
Up to an irrelevant identity contribution, this gives
\begin{eqnarray}
[I_{01},Q_{01}]
\rightarrow
-4i\,\Pi_1,
\qquad
\Pi_1=|1\rangle\langle1|.
\end{eqnarray}
Using the DRAG relation
\begin{eqnarray}
\Omega_Q(t)
=
\frac{\dot\Omega_I(t)}{\Delta},
\end{eqnarray}
and simply using that the integral over $\dot\Omega_I$ reproduces $\Omega_I$, the corresponding first-order BCH (or Magnus) contribution becomes
\begin{eqnarray}
H_{\rm Stark}^{(01)}
=-
\frac{\Omega_I^2(t)}{\Delta}\Pi_1.
\end{eqnarray}

The second contribution originates from the leakage transition operators
\begin{eqnarray}
I_{12}
&=&
|1\rangle\langle2|
+
|2\rangle\langle1|,
\\
Q_{12}
&=&
-i|1\rangle\langle2|
+
i|2\rangle\langle1|.
\end{eqnarray}
Their commutator as arises in both the Schrieffer--Wolff and Magnus expansion treatment is given by
\begin{eqnarray}
[I_{12},Q_{12}]
=
2i
\left(
|1\rangle\langle1|
-
|2\rangle\langle2|
\right).
\end{eqnarray}
Including the matrix-element factor $\lambda_{12}$, this produces the contribution
\begin{eqnarray}
H_{\rm Stark}^{(12,1)}
=
\frac{\lambda_{12}^2\Omega_I^2(t)}{2\Delta}\Pi_1
-
\frac{\lambda_{12}^2\Omega_I^2(t)}{2\Delta}\Pi_2.
\end{eqnarray}

As in the two-level Schrieffer--Wolff derivation of the previous section, the next order in the BCH expansion produces an additional factor of $1/2$.  
This corresponds to the double-commutator correction involving the leakage projector $\Pi_2$ and subtracts half of the previous result,
\begin{eqnarray}
H_{\rm Stark}^{(12,2)}
=
-\frac{\lambda_{12}^2\Omega_I^2(t)}{4\Delta}\Pi_1.
\end{eqnarray}

Combining the three contributions on $\Pi_1$, we finally obtain the effective DRAG detuning correction
\begin{eqnarray}
H_{\rm DRAG}^{(Z)}
=
\frac{\lambda_{12}^2-4}{4\Delta}
\Omega_I^2(t)\Pi_1,
\end{eqnarray}
which can be corrected by appropriate time-dependent detuning. This is combined with the  quadrature correction
$\Omega_Q(t)
=
\frac{\dot\Omega_I(t)}{\Delta}$. In addition, further corrections can be added to reduce errors caused by leakage to even higher order, including multi-photon transitions such as $\ket{0}-\ket{2]}$ and $\ket{1}-\ket{3]}$, see \cite{jesus2025}.

In this section, we have seen some of the most common errors that appear in control theory. We now consider some generalized approaches to deal with arbitrary Hamiltonians.

\section{Adiabatic Methods/Analytic Approaches}
\subsection{Adiabatic Protocols}

Adiabatic protocols exploit the fact that a quantum system prepared in an instantaneous eigenstate of a slowly varying Hamiltonian remains in the corresponding instantaneous eigenstate throughout the evolution. 
For a time-dependent Hamiltonian
\begin{eqnarray}\label{eq:eigenenergies}
H(t)|n(t)\rangle
=
E_n(t)|n(t)\rangle,
\end{eqnarray}
we introduce the unitary transformation
\begin{eqnarray}
U(t)
=
\sum_n |n(t)\rangle\langle n(0)|,
\end{eqnarray}
which maps a fixed reference basis onto the instantaneous eigenbasis of the Hamiltonian. 
Writing the state as
\begin{eqnarray}
|\psi(t)\rangle
=
U(t)|\phi(t)\rangle,
\end{eqnarray}
the Schrödinger equation becomes
\begin{eqnarray}
i\partial_t |\phi(t)\rangle
=
H_{\rm ad}(t)|\phi(t)\rangle,
\end{eqnarray}
with transformed Hamiltonian
\begin{eqnarray}
H_{\rm ad}
=
U^\dagger H U
-
iU^\dagger \dot U.
\end{eqnarray}
The first term is diagonal in the instantaneous eigenbasis, whereas the second term generates couplings between instantaneous eigenstates,
\begin{eqnarray}
-iU^\dagger \dot U
=
-i
\sum_{m,n}
\langle m(t)|\dot n(t)\rangle
|m(0)\rangle\langle n(0)|.
\end{eqnarray}
The off-diagonal matrix elements
\begin{eqnarray}
-i\langle m(t)|\dot n(t)\rangle,
\qquad
m\neq n,
\end{eqnarray}
are the diabatic couplings responsible for nonadiabatic transitions. 
Adiabatic evolution corresponds to the regime where these couplings can be neglected [\cite{kato1950}]. One can formalize this condition by taking the time derivative of Eq.~\eqref{eq:eigenenergies} and multiplying by $\langle m|$ from the left. This yields
 \begin{eqnarray}
\langle m(t)|\dot n(t)\rangle=\frac{\langle m(t)|\dot H | n(t)\rangle}{E_n-E_m},
\qquad
m\neq n.
\end{eqnarray}
We thus find that rate of change of the coupling elements of the Hamiltonian in the instantaneous eigenbasis has to be small compared to the instantaneous level splittings.

In contrast, when these terms are large compared to the change in energy levels over time (such as AC-Stark shifts), these are known as diabatic processes or protocols.

\subsubsection{Landau--Zener dynamics}

The Landau--Zener model describes a driven two-level system with an avoided crossing,
\begin{eqnarray}
H(t)
=
\frac{1}{2}
\left(
vt\,\sigma_z
+
\Delta\,\sigma_x
\right),
\end{eqnarray}
where $v$ is the sweep velocity and $\Delta$ the minimum gap \cite{zener1932}.
Introducing the mixing angle
\begin{eqnarray}
\tan\theta(t)
=
\frac{\Delta}{vt},
\end{eqnarray}
the instantaneous eigenbasis is obtained through the rotation
\begin{eqnarray}
U(t)
=
e^{-i\theta(t)\sigma_y/2}.
\end{eqnarray}
Transforming into the adiabatic basis yields
\begin{eqnarray}
H_{\rm ad}
=
U^\dagger H U
-
iU^\dagger \dot U
=
\frac{\Omega(t)}{2}\sigma_z
-
\frac{\dot\theta(t)}{2}\sigma_y,
\end{eqnarray}
with instantaneous eigenenergies
\begin{eqnarray}
\Omega(t)
=
\sqrt{(vt)^2+\Delta^2}.
\end{eqnarray}

For a sweep from $t=-\infty$ to $+\infty$, we find $|n(\infty)\rangle = U(\infty)U^\dagger (-\infty)|n(-\infty)\rangle$ and thus if we start, e.g., in the initial state $|n(-\infty)\rangle=(1,0)^T$ it will be transferred to $|n(\infty)\rangle=(0,1)^T$.
The exact transition probability is given by the Landau--Zener formula
\begin{eqnarray}
P_{\rm LZ}
=
e^{-\frac{\pi\Delta^2}{2v}}.
\end{eqnarray}
Adiabatic evolution therefore requires sufficiently slow driving compared to the inverse square gap scale. Thus, the method provides an alternative to Rabi oscillation for single-qubit gates, where instead of turning the driving amplitude on and off, chirping the frequency of the qubit provides the same transfer from the $+1$ to $-1$ eigenstate and vice-versa.

In the case of diabatic evolution, $v\gg \Delta^2$, we have the opposite scenario where the start and final eigenstates match, i.e., the transition probability is unity. This corresponds to doing a memory (or identity) gate while changing the operating point of the qubit (e.g., by magnetic field tuning [\cite{Wineland1998TrappedIons, Koch2007Transmon}]).

\subsubsection{STIRAP}

Stimulated Raman adiabatic passage (STIRAP) realizes coherent population transfer in a three-level $\Lambda$-system from state $|1\rangle$ to state $|3\rangle$ while suppressing occupation of the intermediate state $|2\rangle$ [\cite{oreg1984, kuklinski1989, vitanov2017}]. 
Within the rotating-wave approximation, the Hamiltonian is
\begin{eqnarray}
H(t)
=
\frac{1}{2}
\begin{pmatrix}
0 & \Omega_p(t) & 0 \\
\Omega_p(t) & 0 & \Omega_s(t) \\
0 & \Omega_s(t) & 0
\end{pmatrix},
\end{eqnarray}
where $\Omega_p(t)$ and $\Omega_s(t)$ denote the pump and Stokes couplings.

Introducing the mixing angle
\begin{eqnarray}
\tan\theta(t)
=
\frac{\Omega_p(t)}{\Omega_s(t)},
\end{eqnarray}
the instantaneous dark state (where dark refers to the fact that the state $|2\rangle$ that could suffer from spontaneous emission of photons is not populated) is
\begin{eqnarray}
|D(t)\rangle
=
\cos\theta(t)|1\rangle
-
\sin\theta(t)|3\rangle.
\end{eqnarray}
The adiabatic basis is formed by the dark eigenstate $|D(t)\rangle$ and two additional bright eigenstates $|B_{1/2}(t)\rangle$ (with population in $|2\rangle$) of the Hamiltonian. 
Transforming into this basis again yields
\begin{eqnarray}
H_{\rm ad}
=
U^\dagger H U
-
iU^\dagger \dot U.
\end{eqnarray}
The transformed Hamiltonian is diagonal in the adiabatic energies, while the geometric term introduces couplings proportional to $\dot\theta(t)$. 
In particular,
\begin{eqnarray}
-i\langle B_{1/2}(t)|\dot D(t)\rangle
\propto
\dot\theta(t),
\end{eqnarray}
such that adiabaticity requires that $|\dot\theta(t)|$ is small compared to the energy gap, or
\begin{eqnarray}
|\dot\theta(t)|
\ll
\Omega^2_{\rm eff}(t),
\end{eqnarray}
(any combination of slow speed $|\dot\theta(t)|$ and large laser power) with effective gap
\begin{eqnarray}
\Omega_{\rm eff}(t)
=
\sqrt{\Omega_p^2(t)+\Omega_s^2(t)}.
\end{eqnarray}

The characteristic feature of STIRAP is the counter-intuitive pulse ordering, where the Stokes pulse precedes the pump pulse, with $\theta$ going from 0 to $\pi/2$. 
This adiabatically rotates the dark state from $|1\rangle$ to $|3\rangle$ while strongly suppressing population of the lossy intermediate state.

\subsection{Shortcuts to Adiabaticity}

Adiabatic protocols are intrinsically slow since nonadiabatic transitions are suppressed only in the limit of weak diabatic couplings. 
Shortcuts to adiabaticity (STA) are protocols designed to reproduce the final result of adiabatic evolution within finite time (see \cite{Guery-Odelin2019} for a full review). In particular, STA methods are similar to the family of DRAG methods [\cite{Theis2018}] with the main difference that the former ensure faithful state transfer while the latter adiabatically eliminates diabatic leakage subspaces from a resonant (i.e.~diabatic) desired gate transition such as for qudits [\cite{li2025universal}] or entangling qubits [\cite{georgiadis2026,li2024}].
Among the various STA approaches, counterdiabatic (CD) driving provides the most direct geometric interpretation [\cite{demirplak2003, unanyan1997}].

Starting from the transformed Hamiltonian
\begin{eqnarray}
H_{\rm ad}
=
U^\dagger H U
-
iU^\dagger \dot U,
\end{eqnarray}
the second term acts as a gauge potential generated by the time-dependent basis transformation. 
Nonadiabatic transitions arise entirely from its off-diagonal components. 
The idea of counterdiabatic driving is therefore to add an auxiliary Hamiltonian
$H_{\rm CD}(t)$,
such that
\begin{eqnarray}
U^\dagger(H+H_{\rm CD})U
-
iU^\dagger \dot U
\end{eqnarray}
becomes exactly diagonal in the instantaneous eigenbasis.

The exact counterdiabatic Hamiltonian is
\begin{eqnarray}
H_{\rm CD}
=
i
\sum_n
\left(
|\dot n\rangle\langle n|
-
\langle n|\dot n\rangle
|n\rangle\langle n|
\right).
\end{eqnarray}
The second term removes gauge-dependent phases, while the first term exactly cancels the diabatic couplings generated by the moving eigenbasis.

\subsubsection{Counterdiabatic Landau--Zener driving}

For the Landau--Zener Hamiltonian,
the adiabatic-frame Hamiltonian takes the form
\begin{eqnarray}
H_{\rm ad}
=
\frac{\Omega(t)}{2}\sigma_z
-
\frac{\dot\theta(t)}{2}\sigma_y,
\end{eqnarray}
where the second term produces diabatic transitions.
The counterdiabatic Hamiltonian is therefore chosen as
\begin{eqnarray}
H_{\rm CD}(t)
=
\frac{\dot\theta(t)}{2}\sigma_y, \quad
\dot\theta(t)
=
-
\frac{v\Delta}{(vt)^2+\Delta^2},
\end{eqnarray}
and the total Hamiltonian then becomes exactly diagonal in the instantaneous eigenbasis,
\begin{eqnarray}
U^\dagger(H_0+H_{\rm CD})U
-
iU^\dagger \dot U
=
\frac{\Omega(t)}{2}\sigma_z,
\end{eqnarray}
such that the system follows the adiabatic eigenstates exactly even for arbitrarily rapid sweeps. In practice, however, the sweep speed is limited by the available control resources, in particular the maximal admissible pulse amplitude for the counterdiabatic drive which is given by $|\dot\theta(t)|=v/\Delta$.

\subsubsection{Counterdiabatic harmonic oscillator}

Let us consider the harmonic oscillator with time-dependent frequency,
\begin{eqnarray}
H_0(t)
=
\frac{p^2}{2m}
+
\frac{1}{2}m\omega^2(t)x^2.
\end{eqnarray}
Rapid modulation of $\omega(t)$ generally produces excitations between oscillator levels. 
The corresponding adiabatic basis transformation is a squeezing transformation,
\begin{eqnarray}
U(t)
=
\exp\left[
\frac{i}{4}
\ln\frac{\omega(t)}{\omega(0)}\,
(xp+px)
\right].
\end{eqnarray}
Transforming into the instantaneous eigenbasis gives
\begin{eqnarray}
H_{\rm ad}
=
U^\dagger H_0 U
-
iU^\dagger \dot U.
\end{eqnarray}
The geometric contribution evaluates to [\cite{Jarzynski2013,del2013}]
\begin{eqnarray}
-iU^\dagger \dot U
=
\frac{\dot\omega(t)}{4\omega(t)}
(xp+px),
\end{eqnarray}
which generates nonadiabatic squeezing transitions.
The counterdiabatic Hamiltonian 
$H_{\rm CD}(t)
=
iU^\dagger \dot U
$
ensures that
\begin{eqnarray}
U^\dagger(H_0+H_{\rm CD})U
-
iU^\dagger \dot U
\end{eqnarray}
becomes diagonal in the instantaneous oscillator basis. 
The resulting dynamics exactly reproduces the adiabatic evolution of the harmonic oscillator eigenstates in finite time [\cite{muga2010}]. Similar shortcuts can be found for oscillator driving [\cite{jerger2024}], e.g., when simultaneously driving multiple modes.

\subsection{Hamiltonian Engineering}
In Secs.~\ref{sec:blochsiegert} and~\ref{sec:leakage}, we have already seen that the time evolution can in some cases be described by an effective Hamiltonian $H_{\rm eff}$ via some frame transformation or an adequate choice of the control pulse. By slightly changing the perspective, this can also be interpreted as a way to engineer the (effective) Hamiltonian of a system. For example, in Sec.~\ref{sec:leakage}, it has allowed us to engineer a 2-level Hamiltonian from the original 3-level Hamiltonian. 
\subsubsection{Two-qubit gate design and M\o{}lmer-S\o{}rensen gate}

We now consider an important example of Hamiltonian engineering in which qubits are coupled through a common harmonic oscillator mode. 
The basic idea is to apply a spin-dependent force that displaces the oscillator in phase space in a way that depends on the joint qubit state [\cite{Soerensen2000}].

We consider two (or more) qubits (e.g., formed by two electronic levels of a trapped ion) coupled to a single harmonic oscillator with frequency $\omega$ (corresponding to the common vibrational mode in the trap).
Assuming the correct rotating frame, a classical drive near the oscillator sidebands produces
\begin{eqnarray}
H_\text{red} = \eta\Omega\left(\sigma_+ a\, e^{-i\delta t} + \sigma_- a^\dagger e^{i\delta t}\right)
\end{eqnarray}
for the red sideband (where an excitation of the qubit state occurs together with a de-excitation of the harmonic-oscillator state and vice-versa) and
\begin{eqnarray}
H_\text{blue} = \eta\Omega\left(\sigma_+ a^\dagger e^{i\delta t} + \sigma_- a\, e^{-i\delta t}\right)
\end{eqnarray}
for the blue sideband (where the qubit state and the harmonic-oscillator state are always simultaneously (de)-excited), see~\cite{Leibfried2003}.
Driving together $H_\text{red}$ and $H_\text{blue}$ on all qubits simultaneously, produces the interaction
\begin{eqnarray}
H_I(t)
=
\eta\Omega
S_x
\left(
a e^{-i\delta t}
+
a^\dagger e^{i\delta t}
\right),\quad
S_x
=
\sigma_x^{(1)}
+
\sigma_x^{(2)}+\cdots,
\end{eqnarray}
where $S_x$ is the collective spin operator, $\Omega$ denotes the drive strength, $\eta$ the coupling parameter between spin and oscillator, and $\delta$ the detuning from the oscillator sideband.
The interaction therefore acts as a force on the oscillator whose direction depends on the collective spin state. 
To derive the effective qubit interaction, we compute the time evolution generated by $H_I(t)$.

Since the Hamiltonian is linear in $a$ and $a^\dagger$, the first-order Magnus term produces a displacement of the oscillator,
\begin{eqnarray}
-i t \bar H^{(1)}(t)
=
-i\int_0^t dt'\, H_I(t')
=
\alpha(t)a^\dagger S_x
-
\alpha^*(t)aS_x,
\end{eqnarray}
with displacement amplitude
$\alpha(t)
=
\frac{\eta\Omega}{\delta}
\left(
1-e^{i\delta t}
\right)$.
For generic times, the oscillator and the qubits become entangled since different spin states produce different oscillator displacements.
However, the crucial observation is that for gate times satisfying
\begin{eqnarray}
\delta T
=
2\pi n,
\end{eqnarray}
with integer $n$, the displacement vanishes,
$\alpha(T)=0$.
The oscillator therefore returns to its initial state and disentangles from the qubits at the end of the evolution.

A nontrivial qubit interaction nevertheless remains because the Hamiltonian does not commute with itself at different times. 
This effect appears in the second-order Magnus term,
\begin{eqnarray}
-it\bar H^{(2)}(t)
=
-\frac{1}{2}
\int_0^t dt_1
\int_0^{t_1}dt_2\,
[H_I(t_1),H_I(t_2)].
\end{eqnarray}
Using
$[a,a^\dagger]=1$,
the commutator evaluates to
\begin{eqnarray}
[H_I(t_1),H_I(t_2)]
=
-2i
(\eta\Omega)^2
\sin[\delta(t_1-t_2)]
S_x^2.
\end{eqnarray}
The second-order Magnus term therefore becomes
\begin{eqnarray}
-it\bar H^{(2)}(t)
=
i\Phi(t)S_x^2,\quad
\Phi(t)
=
\frac{(\eta\Omega)^2}{\delta^2}
\left(
\delta t-\sin\delta t
\right).
\end{eqnarray}

At the gate time $\delta T=2\pi n$, the sine term vanishes and we obtain the effective evolution
\begin{eqnarray}
U(T)
=
\exp
\left[
i\Phi(T)S_x^2
\right].
\end{eqnarray}
Expanding the collective spin operator as an example for two qubits,
\begin{eqnarray}
S_x^2
=
(\sigma_x^{(1)}+\sigma_x^{(2)})^2
=
2\id
+
2\sigma_x^{(1)}\sigma_x^{(2)},
\end{eqnarray}
we see the generation of entanglement up to an irrelevant global phase.

The M{\o}lmer--S{\o}rensen gate therefore realizes an effective $XX$ qubit--qubit interaction mediated by virtual oscillator excitations. 
The oscillator is only transiently populated during the evolution and returns to its original state at the end of the protocol. Further terms in the Magnus expansion reveal the role of additional dynamical processes, see \cite{Kirchhoff2025}.

\subsubsection{Average Hamiltonian engineering}
A common and elegant approach to Hamiltonian engineering is average Hamiltonian theory. We consider a periodic Hamiltonian $H(t+\tau)=H(t)$ and the time evolution according to the Magnus expansion
\begin{equation}
    U(\tau) = e^{-i\tau (\bar H^{(1)} + \bar H^{(2)}  + \dots)}
\end{equation}
If the Hamiltonian is also symmetric, $H(\tau -t)=H(t)$, one can show that the second-order contribution $\bar H^{(2)}$ vanishes. As a consequence, for $\Vert H(t)\Vert\tau\ll 1$, one can approximate the time evolution by the evolution according to the average Hamiltonian
\begin{equation}
    U(n \tau) \approx e^{-i \bar H^{(1)} n\tau} \,.
\end{equation}
One can now assume that the control allows us to generate some time evolution $U_{\rm frame}(t)$ (e.g., via some pulse sequence) and consider the static Hamiltonian in this so-called toggling frame:
\begin{equation}
    \tilde H (t)= U_{\rm frame}^\dagger (t)H_0 U_{\rm frame}(t)\,,
\end{equation}
where $U(\tau)=\id$ ensures periodicity of $\tilde H(t)$ and the total time evolution then approximately follows the average transformed Hamiltonian
\begin{eqnarray}
    U(n\tau)\approx e^{-in\int_0^\tau \tilde H(t) dt}\,.
\end{eqnarray}

As an example, we consider again the Hahn echo from Sec.~\ref{sec:detuning}. If we apply the protocol twice (one $\pi$-pulse at $\tau/4$ and other one at $3\tau/4$) the pulses generate a toggling frame $U_{\rm frame}(t)$ that is periodic in $\tau$ and with $H_0=\Delta/2\sigma_z$ the transformed Hamiltonian $\tilde H(t)$ is symmetric and the average Hamiltonian vanishes (i.e., the effect of the detuning is decoupled).

With the same pulse sequence, we obtain a slightly more interesting effect if we simultaneously apply it to two qubits with static Hamiltonian
\begin{eqnarray}\label{eq:two-qubit-static}
    H_0=\frac{\Delta_1}{2}\sigma_z\otimes\id + \frac{\Delta_2}{2}\id\otimes\sigma_z + \nu \sigma_z\otimes \sigma_z\,. 
\end{eqnarray}
Such a Hamiltonian describes for example two interacting spins in solid state or two Rydberg atoms in optical traps.
The average transformed Hamiltonian is then simply
\begin{eqnarray}
   \frac{1}{\tau} \int_0^\tau \tilde H(t) dt = \nu \sigma_z\otimes\sigma_z
\end{eqnarray}
and the pulse sequence has then isolated the interaction between the two qubits and decoupled the local detuning terms. This corresponds to a scenario, where, e.g., one wants to entangle the two qubits while the local detuning terms constitute a source of noise.

The opposite effect is desired in some sensing applications, where a magnetic field couples to the local $\sigma_z$ operator, while the interaction between the qubits is disturbing the acquisition of the signal. In the context of NMR, where the qubits are nuclear spins--in this case of the same type -- this is called homonuclear decoupling. A simple sequence for such homonuclear decoupling is the WAHUHA [\cite{Waugh1968}] sequence:
\begin{equation}
    \stackrel{\tau/6}{\longrightarrow}\left(\frac{\pi}{2}\right)_{-Y}
    \stackrel{\tau/6}{\longrightarrow}\left(\frac{\pi}{2}\right)_{X}
    \stackrel{\tau/3}{\longrightarrow}\left(\frac{\pi}{2}\right)_{Y}
    \stackrel{\tau/6}{\longrightarrow}\left(\frac{\pi}{2}\right)_{-X}
    \stackrel{\tau/6}{\longrightarrow}\,.
\end{equation}
For a static Hamiltonian consisting of the local qubit splitting and a dipole-dipole coupling term,
\begin{eqnarray}
    H_0=\frac{\omega}{2}\sigma_z\otimes\id + \frac{\omega}{2}\id\otimes\sigma_z + \nu (2\sigma_z\otimes \sigma_z - \sigma_x\otimes \sigma_x - \sigma_y\otimes \sigma_y)\,,
\end{eqnarray}
this leads to an average Hamiltonian of
\begin{eqnarray}
 \frac{1}{\tau} \int_0^\tau \tilde H(t) dt = \frac{\omega}{6} (\sigma_x+\sigma_y+ \sigma_z)\otimes\id + \frac{\omega}{6} \id\otimes (\sigma_x+\sigma_y+ \sigma_z)\,.
\end{eqnarray}
The result is a precession of the qubits around the effective direction of the qubit splitting, e.g., given by a magnetic field ($\omega=\gamma B_z$).

While we have presented the basic idea, more sophisticated control sequences can generate toggling frames that are robust against pulse errors or additional noise terms in the model, and can isolate or engineer also more general effective Hamiltonians, including with the possibility to tune the coupling coefficients to overcome limitations from the discretization of the time grid to integer multiples of $\tau$, see, e.g. \cite{Vandersypen2005} for NMR systems, \cite{Suter2016} for dynamical decoupling, \cite{Casanova2015} for gate generation and \cite{Choi2020} for sensing protocols.

\subsubsection{Further Reading on Gate Hamiltonian Engineering}
Because of the need to turn entanglement on and off, many such mechanisms exist for engineering two qubit gates through temporary interactions. These fall generally under three categories: (1) engineered couplings through AC fields mediating effective interactions; (2) moving population from the computational space into auxiliary states or mediating elements where interactions occur and then transferring population back again; and (3) bringing qubits to interact with each other either by altering their frequency or spatial separation (e.g. via magnetic fields, flux control, or trapping potentials), thereby increasing the effective interaction strength.

The Mølmer–Sørensen gate is an example of (1), though other prominent mechanisms exist such as the cross-resonance gate for superconducting qubits [\cite{rigetti2010,chow2012,li2024}], and parametric gates based on flux-modulated couplers [\cite{caldwell2018}]. Similar ideas also appear in Raman-mediated gates in NV centers [\cite{childress2006}], as well as exchange-based gates in quantum dots [\cite{burkard1999}], where virtual tunneling through doubly occupied charge states induces an effective Heisenberg interaction. The use of population shelving (2) is especially powerful when combined with strong and long-distance Rydberg–Rydberg blockade interactions [\cite{jaksch2000, Theis2016, jau2016}], as well as long-range cavity-mediated interactions in circuit QED architectures [\cite{blais2007,renger2026}]. Finally, coupling-strength tuning (3) comes into play especially for frequency-tunable superconducting qubits, either for CZ or iSWAP-type interactions via avoided level crossings [\cite{dicarlo2009,strauch2003}], and more generally in collision- or exchange-based gates in neutral atoms [\cite{jaksch1999,calarco2000}] and spin systems [\cite{petta2005}].

\section{Algorithms for Quantum Optimal Control}\label{sec:algorithms}
The core idea of quantum optimal control (QOC) is that a figure of merit (FoM) that quantifies the performance of a control pulse is optimized by varying the control pulse itself. The FoM $J(f)$ thus has to be a function of the control pulse and a control solution is a control pulse $f$ that minimizes (or maximizes) the FoM. There might exist several control solutions and the control solution can then be found exactly or approximately. It can be found via analytic optimization, numerical algorithms or in a closed-loop fashion by evaluating the performance of candidate solutions on an experimental device. In this section we will introduce basic QOC algorithms that are intrinsically relying on gradients (of the FoM with respect to the control pulse) and others that aren't.  Finally, we will discuss tools of machine learning for QOC.

\subsection{Iterative Optimization}
\begin{figure}
    \centering
    \includegraphics[width=\textwidth]{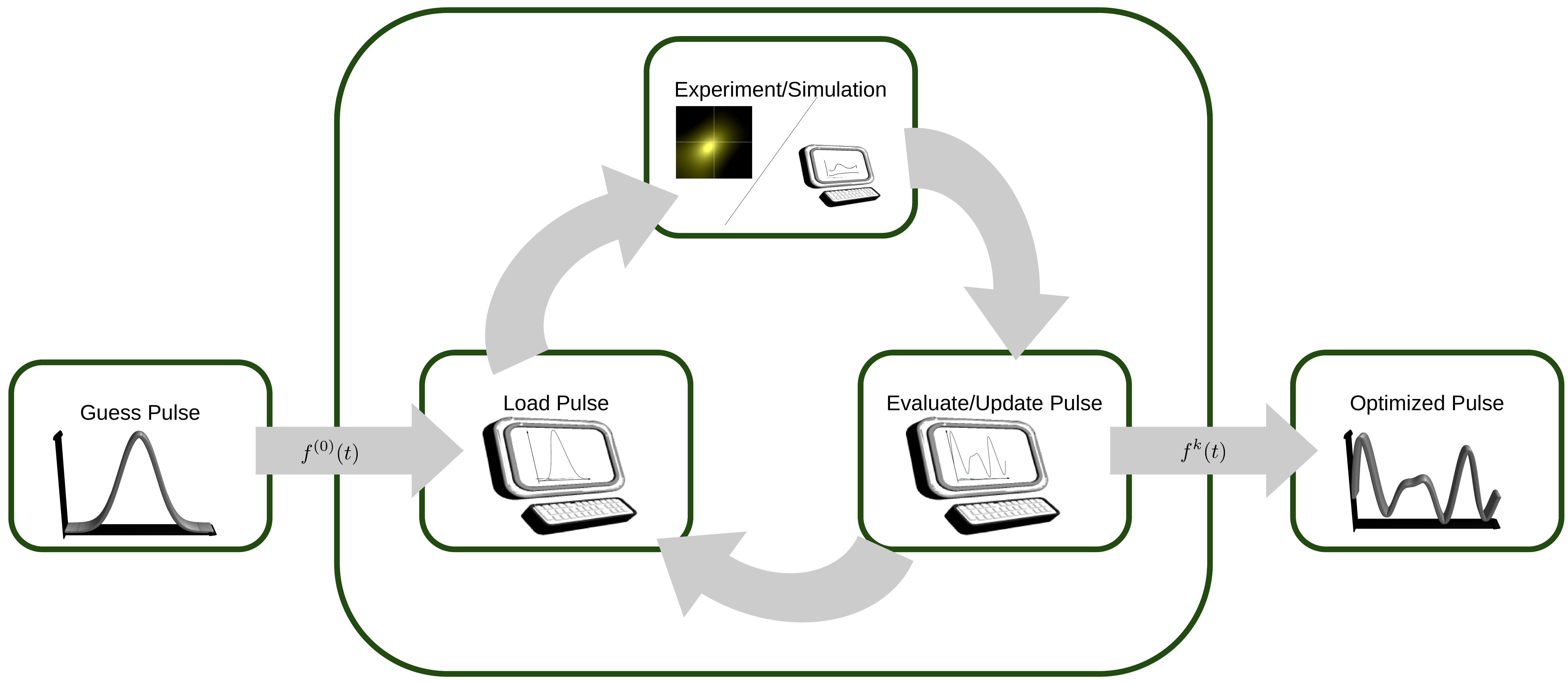}
    \caption{Illustration of Iterative Optimization. Starting from a guess pulse $f^{(0)}(t)$, one iteration of the control algorithm consists of three major steps: The control pulse is loaded on the computer for lab control or into the main simulation for time dynamics. The FoM is then evaluated by performing a measurement on the experimental setup or by performing a numerical simulation on a digital twin. Finally, the control pulse is updated according to the optimization algorithm. These three steps are repeated until the desired performance is reached.}
    \label{fig:iterative-control}
\end{figure}
Most optimization algorithms optimize the pulse step by step. The key procedure is illustrated in Fig.~\ref{fig:iterative-control}. Starting with a guess pulse $f^{(0)}(t)$, the performance of this pulse, i.e., the FoM associated with this pulse or a derived quantity thereof (e.g., the gradient of the FoM with respect to the control pulse) has to be evaluated. Finally, the update algorithm has to update the pulse and deliver the next test pulse $f^{(1)}(t)$. This has to be repeated until after $k$ steps the pulse $f^k(t)$ offers a satisfactory performance. The key difficulty lies in the evaluation of the pulse performance. This can be done numerically by performing a simulation of the time evolution of the system and then extracting all the desired quantities regarding the FoM. In this case a very faithful numerical description of the quantum device (including all the control electronics, \dots) is needed. Where this is not feasible (as for example when such a faithful model is not known, some system parameters can't be determined with sufficient precision, or the simulation is too time-consuming), the evaluation of the FoM can also be done by applying the test pulse to the experiment and by performing measurements on the system. One can distinguish these two approaches as open-loop (the optimal pulse is prepared on the computer and then sent to the device only at the very end) and closed-loop (the feedback from the experiment is used to further optimize the pulse) optimization. While in open-loop optimization the modeling and numerics are challenging, in closed-loop optimization, the evaluation of the FoM through a noisy readout process poses the greatest challenge. In this regard, methods like gate-set tomography or randomized benchmarking can help to overcome limitations from the readout error and symmetries in the time-dynamics can help to reduce the degrees of freedom incorporated in the FoM and thus the complexity of the control problem.

\subsection{Gradient-based Optimization}

The FoM can be for example the residual error of a state transfer
\begin{eqnarray}
	J=1-F\quad \text{with}\quad F=|\langle \psi_G |\psi_f(T)\rangle|^2\,,
\end{eqnarray}
but in more general terms it is simply a (smooth) function of the control pulse via some dynamical equation (e.g., the time evolution relating the initial state to the final state), a way to quantify the performance (in this case the overlap of final state and target state) and potentially some additional constraints (e.g., the maximal admissible pulse amplitude, the bandwidth of the control pulse, \dots). The main idea of gradient algorithms can be understood by looking at the change of the FoM for a small pulse update $\Delta f$:

\begin{eqnarray}
	J(f+\Delta f)\approx J(f)+\frac{\dd J}{\dd f} \Delta f  + \frac{1}{2}\frac{\dd^2 J}{\dd  f ^2}\Delta f^2\,.
\end{eqnarray}
For a small ($\varepsilon>0$) pulse update in the negative direction of the gradient $\Delta f = - \varepsilon \frac{\dd J}{\dd f}$ (neglecting the second order contribution) we get
\begin{eqnarray}
	J(f+\Delta f)\approx J(f)+ \frac{\dd J}{\dd f} \Delta f  = J(f) -\varepsilon \left(\frac{\dd J}{\dd f}\right)^2 < J(f)\,.
\end{eqnarray}
As a consequence, we can minimize the FoM iteratively by moving down along the gradient.

If we have a closer look at the gradient $\frac{\dd J}{\dd f}$ we notice that we have so far ignored that $f(t)$ is typically a time-dependent function, that we might have more than one control, that the step size $\varepsilon$ is a free parameter of the algorithm and could be chosen in different ways, that there might be different ways to build in constraints on the control pulse and that also second-order terms in the expansion of $J$ (or Hessian) can be used. This leads to many different gradient-based algorithms and here we want to describe in more detail the GRAPE (GRadient-Ascent-Pulse-Engineering) and the Krotov algorithm.

\subsubsection{GRAPE}
The GRAPE algorithm [GRadient Ascent Pulse Engineering, \cite{Khaneja2005}] considers the time-dependent control pulse on a time grid (often corresponding to the time grid of the numerical time evolution) and we have to thus discretize the pulse by assigning a value $f_k$ for every time step $t_k$:
\begin{eqnarray}
	f(t)\longrightarrow f_k=f(t_k)\,.
\end{eqnarray}
If we assume that each grid point represents a time interval $\Delta t$ and the time evolution is governed by a Hamiltonian $H(f(t))$, we can discretize also the (unitary) time evolution. For the interval $k$ we obtain
\begin{eqnarray}
	U_k=e^{-i H(f_k)\Delta t}\stackrel{\text{typically}}{=}e^{-i (H_0 + f_k H_c)\Delta t}\,.
\end{eqnarray}
The total time evolution and the time evolution from $t_0$ to $t_k$ are then
\begin{eqnarray}
	U(T)=\mathcal{T} \prod_{k=1}^{N_t}U_k\\
	U(t_k)=\mathcal{T} \label{eq:grapeprop}\prod_{j=1}^{k}U_j\,,
\end{eqnarray}
where the time-order operator simply means that unitaries with higher index have to be multiplied from the left. The gradient of the local time evolution $U_k$ with respect to a change of $f(t)$ can then be approximated as
\begin{eqnarray}
	\frac{\partial U_k}{\partial f_j}=-i\delta_{jk} H_c \Delta t U_k\,.
\end{eqnarray}
We thus see that (by construction) the local gradient $\frac{\partial U_k}{\partial f_j}$ just depends on the value of the control pulse $f_k$ in the same time interval. Note that in practice this will only work for small $\Delta t$ and most implementations of GRAPE use more sophisticated eigenvalue decomposition to calculate the gradient [\cite{Defouquieres2011,motzoi2011}], meanwhile including curvature information in the form of the Hessian [\cite{dalgaard2020}] or its approximation [\cite{Defouquieres2011}]. This dramatically decreases the number of steps required, often finding very high fidelity solutions with only a few dozen iterations.

If we go back to our state-transfer problem $J=1-F$, we can find explicit expressions for the update rules. First, the final state depends on the control pulse $f$ via the time evolution
\begin{eqnarray}
 |\psi_f(T)\rangle=U(T)|\psi_0\rangle\,.
\end{eqnarray}
The gradient of the FoM can then be calculated as
\begin{eqnarray}
	\frac{\dd J}{\dd f}=-\frac{\dd F}{\dd f} = -2\re \langle\psi_f(T)|\psi_G\rangle \left\langle \psi_G\left| \frac{\dd}{\dd f}\right|\psi_f(T)\right\rangle\,.
\end{eqnarray}
If we vary the pulse only in interval $k$, i.e., we calculate the gradient of $J$ with respect to $f_k$, only $U_k$ will change:
\begin{eqnarray}
	\frac{\dd J}{\dd f_k}= -2\re \langle\psi_{f}(T)|\psi_G\rangle\left\langle \psi_G\left|U_{N_t} U_{N_t-1}\dots U_{k+1} \frac{\partial U_k}{\partial f_k}U_{k-1}\dots U_1\right|\psi_0 \right\rangle\,.
\end{eqnarray}
We can introduce the so-called forward-propagated and backward-propagated states
\begin{eqnarray}\label{eq:propstates}
	|\psi_{fw}(t_k)\rangle=U_{k}\dots U_1|\psi_0\rangle\\
 \label{eq:propstates2}
	|\psi_{bw}(t_k)\rangle=U_{k+1}^\dagger\dots U_{N_t}^\dagger|\psi_G\rangle\,.
\end{eqnarray}
In this way, both the forward-propagated and the backward-propagated states fulfill the same discretized Schrödinger equation, but with different initial conditions: $|\psi_{fw}(0)\rangle=|\psi_0\rangle$ and $|\psi_{bw}(T)\rangle=|\psi_G\rangle$.
Substituting into the formula for the gradient the expression simplifies to
\begin{eqnarray}
	\frac{\dd J}{\dd f_k}= -2\re \langle\psi_{fw}(T)|\psi_G\rangle \left\langle \psi_{bw}(t_{k})\left|\frac{\partial U_k}{\partial f_k}\right|\psi_{fw}(t_{k-1})\right\rangle.
\end{eqnarray}
The basic version of GRAPE now consists in iteratively sweeping through the time interval and updating the control pulse at each time step according to the update formula
\begin{eqnarray}\label{eq:update_GRAPE}
	f_k \rightarrow f_k - \varepsilon \frac{\dd J}{\dd f_k}, 
\end{eqnarray}
where again $\frac{\partial U_k}{\partial f_j}$ can be calculated more precisely using eigendecomposition and/or Hessian information. Similar update rules can be found for gate optimization, open system dynamics, and a host of other problems. The main advantage of the GRAPE algorithm compared to naive gradient descent comes from Eqs.~\eqref{eq:propstates}-\eqref{eq:propstates2} being stored and subsequently reused in both the backwards and forwards passes, eliminating the need for multiple recalculations. This mechanism closely resembles the speedup from backpropagation in neural network parameter updates, though nowadays these types of algorithmic speedups are implicitly performed by using automatic differentiation methods for gradient calculations.

\subsubsection{Krotov}
The Krotov algorithm [\cite{Konnov1999}] follows a very similar update formula and we will use this section to present an explicit form for more general control problems than state transfer. For the derivation of the Krotov algorithm we approach the problem in a slightly different way. We first add a constraint to the FoM
\begin{eqnarray}
	J=-F + \int_0^T L(f^{k+1}(t))\dd t
\end{eqnarray}
that limits the difference between the old pulse $ f^{k}(t)$ and the updated pulse $f^{k+1}(t)$, via
\begin{eqnarray}
	L(f^{k+1}(t))=\lambda(t)(f^{k+1}(t)-f^{k}(t))^2\,.
\end{eqnarray}
Let's assume we have $N$ basis states $\{|n\rangle\}_{n=1,\dots N}$ with corresponding forward-propagated states $\{|\psi_{fw,n}^{k}(t)\rangle\}_{n=1,\dots N}$ ($|\psi_{fw,n}^{k}(0)\rangle=|n\rangle$) where $k$ refers to the control pulse after $k$ iterations, and $F$ depends on the value of these states for the next iteration $k+1$ at final time $|\psi_{fw,n}^{k+1}(T)\rangle$. In the same way we define $N$ backward-propagated states $\{|\psi_{bw,n}^{k}(t)\rangle\}_{n=1,\dots N}$ that fulfill the Schrödinger equation and $|\psi_{bw,n}^{k}(T)\rangle=|\tilde n\rangle$, where also $\{|\tilde n\rangle\}_{n=1,\dots N}$ is a basis and not necessarily $|n\rangle =|\tilde n\rangle$.
The FoM can now be interpreted as action integral and the minimization of the action integral requires
\begin{eqnarray}
	\frac{\dd J}{\dd f}=\frac{\partial J}{\partial f}+\partial_n J \left(\frac{\dd \psi_{fw,n}^{k+1}(T)}{\dd f}\right)=0\,.
\end{eqnarray}
Here, $\pt_n J$ means the Fréchet derivative of $J$ with respect to the $n$-th argument (i.e., the state $\psi_{fw,n}$, which is a function of time).
The challenge arises now from the fact that $L$ depends on the control pulse $f^{k+1}(t)$ in an explicit way while $F$ depends on $f^{k+1}(t)$ in an implicit way via the time-evolution of the states. We can try to separate the different contributions to the total derivative of the FoM with respect to the control pulse by introducing an additional functional $\phi$ via
\begin{eqnarray}
	J = J - \phi(0,\{|n\rangle\}) + \phi(T,\{\psi_{fw,n}^{k+1}(T) \}) - \int_0^T \frac{\dd}{\dd t}\phi(t,\{\psi_{fw,n}^{k+1}(t) 
	\}) \dd t 
\end{eqnarray}
and demanding two more rigid conditions for minimizing $J$:
\begin{eqnarray}
	\frac{\partial J}{\partial f}=0\\
	\partial_n J =0\,.
\end{eqnarray}
By introducing
\begin{eqnarray}
	G&=&-F + \phi(T,\{ \psi_{fw,n}^{k+1}(T) \})\\
	R&=& -L(f^{k+1}(t)) + \frac{\pt}{\pt t}\phi(t,\{ \psi_{fw,n}^{k+1}(t) \}) \nonumber\\
	&&+ \sum_{n=1}^N  \pt_{n}\phi(t,\{ \psi_{fw,n}^{k+1}(t) \})(-i H(f^{k+1}(t)) \psi_{fw,n}^{k+1}(t))
\end{eqnarray}
we bring the FoM to the form
\begin{eqnarray}
	J=G-\phi(0,\{|n\rangle\}) - \int_0^T R\dd t\,.
\end{eqnarray}
We can make a quadratic ansatz (often a linear ansatz, i.e., $\sigma(t)=0$ is sufficient; see discussion below)
\begin{align}
	\phi(t,\{\psi_{fw,n}^{k+1}\})=\sum_{m,n=1}^{N}\re c_{mn}\langle \psi_{bw,m}^{k}(t)\vert \psi_{fw,n}^{k+1}(t)\rangle + \nonumber\\
	+ \frac {1}{2}\sum_{n=1}^{N} \sigma(t) \langle \psi_{fw,n}^{k+1}(t)-\psi_{fw,n}^{k}(t)\vert \psi_{fw,n}^{k+1}(t)-\psi_{fw,n}^{k}(t) 
	\rangle\,.
\end{align}
The functional $\phi$ thus depends explicitly on time (via the old states $|\psi_{fw,n}^{k}(t)\rangle$ and $|\psi_{bw,n}^k(t)\rangle$ as well as via $\sigma(t)$) and on the states $\{\psi_{fw,n}^{k+1}\}$.
By this choice of $\phi$ and $R$, the condition $\partial_n J= 0$ reduces to $\partial_n G=0$  (one can check that $\partial_n R=0$) which allows us to determine the coefficients $c_{mn}$:
\begin{eqnarray}
	\pt_{n}\phi(t,\{\psi_{fw,l}^{k}(T)\})(\varphi)=\sum_{m=1}^{N}\re c_{mn}\langle\psi_{bw,m}^{k}(T)\vert 
	\varphi\rangle \stackrel{!}{=}\pt_{n}F(\{\psi_{fw,l}^{k}(T)\})(\varphi)\;\forall\varphi\,.
\end{eqnarray}
Finally, the explicit dependence of the FoM on the control pulse allows us to derive the update formula for the Krotov algorithm. From $\pt J/\pt f = \pt R/\pt f=0$ we find
\begin{align}\label{eq:Krotov-update-general}
	f^{k+1}(t)=&f^k(t)+\frac 1{2\lambda(t)}\Bigg [\sum_{m,n=1}^N \re \, c_{mn}\Big\langle\psi_{bw,m}^k(t)\Big|-i\frac{\pt H}{\pt f}(f^{k+1}(t))\Big|\psi_{fw,n}^{k+1}(t)\Big\rangle\nonumber\\
	&+\sum_{n=1}^N\sigma(t)\re\Big\langle\psi_{fw,n}^{k+1}(t)-\psi_{fw,n}^k(t)\Big|-i \frac{\pt H}{\pt f}(f^{k+1}(t))\Big|\psi_{fw,n}^{k+1}(t)\Big\rangle\Bigg]\,.
\end{align}
The two constants $\lambda(t)$ and $\sigma(t)$ control the step-size of the algorithm and the quadratic correction. Too large step-sizes (too small $\lambda(t)$) can cause the FoM to increase instead of decreasing because of the nonlinear dependence of $J$ on the control pulse. The quadratic correction instead is helping to ensure monotonic convergence of the FoM even when the FoM (typically via $F$) depends nonlinearly on the states (then $\sigma(t)$ should take on a sufficiently large negative value, see~\cite{Sklarz2002,Reich2012}) and is not required (i.e., one can choose $\sigma=0$) in the case of standard state transfer or gate optimization.
An implementation of the algorithm is available in the Krotov package by \cite{Goerz2019}.

\begin{figure}
	\centering
	\includegraphics[width=0.45\linewidth]{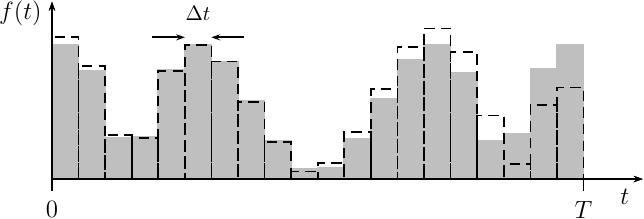}
	\includegraphics[width=0.45\linewidth]{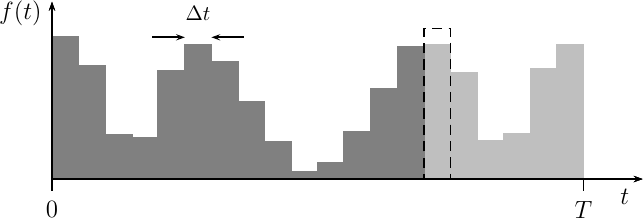}
	\caption{Pulse update for GRAPE and Krotov. The left panel shows the concurrent pulse update for GRAPE. The old pulse (grey) is updated for all time steps all at once to the new pulse (black dashed lines). The right panel, instead, shows the pulse update for Krotov, where the pulse update is sequential. The new pulse is shown in dark grey where it is already updated. This is used for the forward-propagated states. The backward-propagated states, instead are calculated by the old pulse (light grey) and the pulse is updated only for one time slice after the other.}
	\label{fig:pulse_update_gradient}
\end{figure}

We can now compare the update formula for the Krotov algorithm with the update formula for GRAPE by bringing  Eq.~\eqref{eq:update_GRAPE} to a form similar to Eq.~\eqref{eq:Krotov-update-general}:
\begin{align}
	f^{k+1}(t)=&f^k(t)+\frac 1{2\lambda(t)}\sum_{m,n=1}^N \re \, c_{mn}\Big\langle\psi_{bw,m}^k(t)\Big|-i\frac{\pt H}{\pt f}(f^{k}(t))\Big|\psi_{fw,n}^{k}(t)\Big\rangle\,.
\end{align}
We can see that this is the linear version of the Krotov update with the key difference that both the backward-propagated and the forward-propagated pulse are calculated with the old pulse. This means that for Krotov the pulse is updated sequentially for each time step taking into account already the update for the previous time step, while for GRAPE the update is concurrent, i.e., the pulse is updated at once for all time steps. The concept is illustrated in Fig.~\ref{fig:pulse_update_gradient}.

We now consider the explicit form of the pulse update for the examples of state transfer and gate optimization.
For the state transfer, we have $N=1$ and can thus skip the indices referring to the basis state and set $c_{mn}=c$, $\ket{\psi_{fw/bw,n}^{k}}=\ket{\psi_{fw/bw}^{k}}$. The backward-propagated state starts from the target state $\ket{\psi_{bw}^{k}(T)}=\ket{\psi_G}$ and the forward-propagated state from the initial state $\ket{\psi_{fw}^{k}(0)}=\ket{\psi_0}$. The state-transfer fidelity is $F=|\langle\psi_G|\psi_{fw}(T)\rangle|^2$ and the condition on the indices is
\begin{align}
	\partial F(\varphi)&=2\mathrm{Re}\langle\psi_{fw}^k(T)|\psi_G\rangle\langle\psi_G|\varphi\rangle\nonumber\\
	&\stackrel{!}{=}\mathrm{Re}\,c\,\langle\psi_{bw}^k(T)|\varphi\rangle=\mathrm{Re}\,c\,\langle\psi_G|\varphi\rangle
\end{align}
and thus $c=2\langle\psi_{fw}^k(T)|\psi_G\rangle$ which is consistent with Eq.~\eqref{eq:update_GRAPE} for $1/\lambda(t)=2\varepsilon\Delta t$ and $\frac{\pt H}{\pt f}=H_c$.

For gate optimization, here, we consider the gate fidelity $F_{sm}$
\begin{align}
	F=F_{sm}=\frac 1 {N^2}\left|\sum_{n=1}^N\langle n|V^\dagger|\psi_{fw,n}(T)\rangle\right|^2\,.
\end{align}
Defining $v_{mn}=\langle n |V|m\rangle$ we obtain $V\ket{m}=\sum_{n=1}^N v_{mn}| n\rangle$.
Choosing $\ket{\psi_{bw,n}^{k}(T)}=\ket{n}$ (i.e. $|n\rangle=|\tilde n\rangle$) we find
\begin{align}
    F&=\frac 1 {N^2}\sum_{m,l=1}^N\langle m|V^\dagger|\psi_{fw,m}(T)\rangle\langle \psi_{fw,l}(T)|V|l\rangle\nonumber\\
	\partial_n F(\varphi)&=\frac 2 {N^2}\re\sum_{l=1}^N \langle\psi_{fw,l}(T)|V|l\rangle \langle n|V^\dagger |\varphi\rangle=\frac 2 {N^2}\re\sum_{m,l=1}^N \langle\psi_{fw,l}(T)|V|l\rangle v_{nm}^{*}\langle m|\varphi\rangle\nonumber\\
	&\stackrel{!}{=}\sum_{m=1}^N\re\,c_{mn}\langle\psi_{bw,m}^k(T)|\varphi\rangle=\sum_{m=1}^N\re\,c_{mn}\langle m|\varphi\rangle\,.
\end{align}
and thus $c_{mn}=\frac 2 {N^2}\sum_{l=1}^N \langle\psi_{fw,l}(T)|V|l\rangle v_{nm}^{*}$.

\subsection{Gradient-free Optimization}
Gradient-free algorithms are based on an expansion of the control pulse into a basis. A trivial basis is given by the discretization of the time as in the case of the direct gradient-based algorithms. However, the efficiency of the optimization requires one to limit the amount of parameters or basis functions and so a piece-wise constant control pulse with only a limited number of steps might not always be desirable. Instead, smooth basis functions are typically preferred, such as the (random) Fourier basis or step functions with a smooth step such as the sigmoid basis [\cite{Pagano2024}]. Once parameterized, the coefficients of the expansion can be optimized by any standard optimization algorithm, such as Nelder-Mead, CMA-ES, Bayesian optimization or Monte-Carlo algorithms. Note that arbitrary basis functions and gradient descent (typically via GRAPE) are not mutually exclusive and they can be readily combined, see \cite{motzoi2011, Soerensen2018}. In particular, one can perform the direct optimization of the coefficients by exploiting the gradient obtained by automatic differentiation [\cite{Rossignolo2023}]. Here, we take a closer look at an efficient way to choose the basis, the Chopped RAndom Basis (CRAB) algorithm [\cite{Doria2011,Caneva2011}] for gradient-free QOC.

\subsubsection{CRAB and dCRAB}
We start by an expansion of the control pulse in a basis with $N_c$ (CRAB) basis functions $f_i(t)$:
\begin{eqnarray}
	f(t)=\sum_{i=1}^{N_c} c_i f_i(t)\,,
\end{eqnarray}
where $c_i$ are the coefficients that have to be optimized in order to minimize the FoM. As an example for the choice of basis functions we consider the trigonometric functions
\begin{eqnarray}
	f_i(t)=\sin(\omega_i t)\quad\text{or}\quad \cos(\omega_i t)
\end{eqnarray}
which are often used together
\begin{eqnarray}
	f(t)=\sum_{i=1}^{N_c} a_i \cos(\omega_i t) + b_i \sin(\omega_i t)\,.
\end{eqnarray}
In principle, any set of basis functions can be used for CRAB, however, it has been proven advantageous to include a random element in this choice of basis functions so that different random instances of the truncated control space can be explored. In the case of the trigonometric basis functions, this random element are the frequencies $\omega_i$ which are chosen randomly around the harmonics thus varying slightly the standard Fourier basis.

To make the resulting control pulse even smoother, especially at initial and final times, one can introduce an envelope function $\lambda_C(t)$ via
\begin{eqnarray}
	f(t)=\lambda_C(t)\sum_{i=1}^{N_c} c_i f_i(t)\,.
\end{eqnarray}
This envelope function can be for example a Gaussian or it can have a $\sin^n$-shape over a certain rise time $\tau$ during the time intervals [$0,\tau$] and [$T-\tau,T$]. Additionally, if a rough solution is already known (e.g., from analytical considerations or a previous partial optimization attempt), one can use this rough solution as a guess pulse $g(t)$:
\begin{eqnarray}
	f(t)=g(t) + \lambda_C(t)\sum_{i=1}^{N_c} c_i f_i(t)\,.
\end{eqnarray}
A similar update formula can also be used to introduce additional iterations of the algorithm with new basis functions, so-called super-iterations. We label by $j$ the super-iteration and consider
\begin{eqnarray}
	f^{(j)}(t)=f^{(j-1)}(t) + \lambda_C(t)\sum_{i=1}^{N_c} c^{(j)}_i f^{(j)}_i(t)\,.
\end{eqnarray}
Then, $f^{(j-1)}(t)$ is the control solution obtained by the previous super-iteration, $f^{(j)}_i(t)$ are the basis functions for super-iteration $j$, and $f^{(j)}(t)$ is the control pulse that has to be optimized in super-iteration $j$ by optimizing the coefficients $c^{(j)}_i $. This extension of CRAB to super-iterations is called dressed CRAB (dCRAB) algorithm and is capable of removing the local minima in the control landscape introduced by the basis expansion [\cite{Rach2015}].
 
For further reading, we refer to the review by \cite{Mueller2022}. An implementation is available in the QuOCS package by \cite{Rossignolo2023}.

\subsection{Learning-Based Quantum Control}
A natural way to formulate learning-based quantum control is to view a neural network as a parameterized map between physically meaningful inputs and outputs. Two especially natural formulations are learning families of optimal pulses and learning the fidelity landscape itself.
\subsubsection{Learning families of optimal pulses}

Suppose the Hamiltonian depends on parameters $\lambda$ (detuning, coupling strengths, noise amplitudes, target gate parameters, etc.), and the goal is to predict an optimal control pulse $u(t)$. One may write
\begin{equation}
H(t)=H_0(\lambda)+u(t)H_c ,
\end{equation}
and approximate the control policy with a neural network
\begin{equation}
u_\theta(t)=f_\theta(\lambda,t),
\end{equation}
where $\theta$ are trainable weights. Training minimizes a fidelity-based loss,
\begin{equation}
\mathcal{L}(\theta)=1-F\!\left(U_\theta(T),U_{\mathrm{target}}\right).
\end{equation}

This presentation makes the neural network resemble a surrogate optimal-control solver: given Hamiltonian or gate parameters, it directly outputs a pulse sequence. Such approaches can learn continuous families of controls and generalize across related optimization tasks, reducing the need for repeated pulse optimization from scratch. Examples include \cite{preti2022} and \cite{sauvage2022}.

\subsubsection{Learning the fidelity landscape}

Alternatively, the neural network can approximate the mapping from pulse parameters to achieved fidelity [\cite{dalgaard2022, calzavara2025}]. Let the pulse be parameterized by coefficients $u=(u_1,\dots,u_n)$. Then
\begin{equation}
F_\theta(u)\approx F_{\mathrm{true}}(u),
\end{equation}
with training data $(u^{(i)},F_i)$ and training objective
\begin{equation}
\mathcal{L}(\theta)=\frac{1}{N}\sum_i
\left[F_\theta(u^{(i)})-F_i\right]^2 .
\end{equation}

In this view, the neural network becomes a fast differentiable surrogate model of the control landscape, enabling accelerated optimization and exploration of high-dimensional pulse spaces. Such surrogate fidelity models also form a natural bridge to reinforcement learning, where neural networks similarly approximate value functions such as $Q_\theta(s,a)$ or expected future rewards.

\subsection{Reinforcement Learning}

Quantum control seeks a sequence of control pulses that drives a quantum system toward a desired target state or gate while respecting experimental constraints. 
In the QOC algorithms above the idea is to iteratively improve the FoM, e.g., moving along the gradient or in the vicinity of the currently best pulse. 
Reinforcement learning (RL) instead frames control as a sequential decision problem: an agent interacts with a quantum environment, applies control actions, and receives rewards based on the achieved fidelity.

In RL, the agent observes a state $s_t$, selects an action $a_t$, and receives reward $r_t$. 
The goal is to maximize the expected cumulative return
\begin{eqnarray}
G_t
=
\sum_{k=0}^{\infty}
\gamma^k r_{t+k},
\end{eqnarray}
where $\gamma\in[0,1]$ is the discount factor suppressing future rewards. 
The central quantity is the action-value function
\begin{eqnarray}
Q^\pi(s,a)
=
\mathbb{E}_\pi
\left[
G_t
\,\big|\,
s_t=s,
a_t=a
\right],
\end{eqnarray}
where $\pi$ is known as the policy, which is the chosen method or protocol for choosing a next action for a given state [\cite{sutton2018}].
Using the recursive structure
\begin{eqnarray}
G_t
=
r_t
+
\gamma G_{t+1},
\end{eqnarray}
one obtains the Bellman equation
\begin{eqnarray}
Q^\pi(s,a)
=
\mathbb{E}
\left[
r
+
\gamma
Q^\pi(s',a')
\right].
\end{eqnarray}
This recursive relation underlies most RL algorithms used in quantum control, allowing future actions under many trials (known as episodes) to reinforce the best policy for the present action.

In practice, the quantum state, measurement outcomes, or reduced observables define the state $s_t$, while the action $a_t$ corresponds to pulse amplitudes, frequencies, phases, or pulse durations. 
The reward is typically related to gate fidelity, state overlap, leakage suppression, or energetic cost.

\subsubsection{Classical reinforcement learning: Q-learning}

A simple off-policy RL method is Q-learning, which updates the action-value function according to
\begin{eqnarray}
Q(s,a)
\leftarrow
Q(s,a)
+
\alpha
\left[
r
+
\gamma
\max_{a'}
Q(s',a')
-
Q(s,a)
\right],
\end{eqnarray}
where $\alpha$ denotes the learning rate.

Because the update uses the maximizing action independently of the current behavior policy, Q-learning is an \textit{off-policy} algorithm [\cite{watkins1992}]. 
For low-dimensional quantum systems, the state-action space may remain sufficiently small that tabular RL can directly discover effective pulse sequences [\cite{bukov2018}]. 
Such methods are computationally lightweight and interpretable, but they scale poorly with increasing Hilbert-space dimension due to the exponential growth of the quantum state space.

\subsubsection{Deep reinforcement learning: PPO, DDPG and AlphaZero}

For realistic quantum devices, the state and action space become continuous and exponentially large. 
Deep reinforcement learning therefore replaces the tabular value function with a neural-network approximation.

A widely used class of deep reinforcement-learning algorithms for continuous
action spaces includes proximal policy optimization [PPO, \cite{schulman2017}] and deep deterministic
policy gradient [DDPG, \cite{lillicrap2015}] in the context of quantum control [\cite{an2021, nam2024, ernst2025}]. These methods optimize parameterized neural-network
policies
\begin{eqnarray}
\pi_\theta(a|s),
\end{eqnarray}
where $\theta$ denotes the neural-network parameters that determine the relation between the current state and the prescribed action. 
PPO employs stochastic policy
updates constrained for training stability, whereas DDPG learns a deterministic
continuous-action policy together with a learned value-function approximation
in an off-policy framework.

In quantum control, deep reinforcement-learning agents can learn adaptive pulse strategies that are robust against decoherence, calibration drift, and measurement noise. 
Neural-network policies are particularly useful when the environment is only partially observable or when rewards are obtained only after completion of an entire pulse sequence.

Another class of approaches is inspired by AlphaZero-style algorithms \cite{silver2018}, which combine value estimation via Monte Carlo tree search with deep neural networks [\cite{dalgaard2020b}]. 
The search procedure explores candidate control trajectories while the neural network estimates long-term value functions and promising actions. 
Compared with PPO, such methods are computationally more demanding but can provide improved planning performance in highly structured control landscapes.

A useful distinction between these methods is therefore:
\begin{itemize}
\item Off-policy methods (e.g. Q-learning, DDPG): sample efficient, reusable experience, simple updates.
\item On-policy methods (e.g.~PPO): more stable for continuous control and neural-network policies.
\item Search-enhanced methods (e.g.~AlphaZero): combine planning and learning for high-dimensional control problems.
\end{itemize}

Reinforcement learning has become increasingly attractive for quantum technologies because it combines exploration with long-horizon value estimation, allowing agents to identify promising control trajectories even when intermediate feedback is incomplete or noisy. 
Current research increasingly incorporates RL with physics-informed neural-network architectures and differentiable quantum simulators in order to improve both sample efficiency and physical interpretability [\cite{norambuena2024, khalid2023}].

\subsection{Constraints}
The control algorithms can be varied by including additional constraints. These constraints typically are constraints on the control pulse, but can be also constraints on a certain trajectory in the Hilbert space or a subspace. We have already seen that a bandwidth constraint naturally arises from the expansion of a control pulse into a bandwidth-constrained basis. Alternatively, it could be introduced by combining the gradient-based updates with frequency filters [see \cite{motzoi2011} and \cite{singh2023} for linear and nonlinear transfer functions, respectively]. Additionally, amplitude or energy constraints are very common. These can be introduced by changing the update formula or via Lagrange multipliers or penalty terms. In this second case, one can for example introduce a penalty term on the maximum value of the control pulse
\begin{eqnarray}\label{eq:pulseheight-weight}
	J=1-F+\lambda \max_{t}|f(t)|\,,
\end{eqnarray}
or on its energy
\begin{eqnarray}\label{eq:pulsepower-weight}
	J=1-F+\lambda \int_0^T f(t)^2 dt\,.
\end{eqnarray}
Alternatively, such a constraint on the pulse amplitude could also be directly built into the update formula. For CRAB, e.g., the pulse update can be modified by 
\begin{eqnarray}\label{eq:pulseheight-wall}
	\tilde{f}(t)=\lambda(t)\sum_{i=1}^{N_c}c_i f_i(t)\,,\\
	f(t)=\begin{cases}
		\tilde{f}(t)&\text{if}\,\,|\tilde{f}(t)|<f_{\mathrm{max}}\\
		\mathrm{sgn}(\tilde f(t)) f_{\mathrm{max}}&\text{otherwise},
	\end{cases}
\end{eqnarray}
or in a smoother version
\begin{eqnarray}\label{eq:pulseheight-wall2}
	\tilde{f}_{\mathrm{max}}=\max_{t} |\tilde{f}(t)|\,,\\
	f(t)=\begin{cases}
		\tilde{f}(t)&\text{if}\,\,\tilde{f}_{\mathrm{max}}<f_{\mathrm{max}}\\
		\frac{f_{\mathrm{max}}}{\tilde{f}_{\mathrm{max}}}\tilde{f}(t) &\text{otherwise}.
	\end{cases}
\end{eqnarray}
While the first modification of the update formula simply truncates the control pulse at the maximal admissible amplitude (and thus introduces discontinuities of its derivative), the second approach maintains the bandwidth constraints introduced by the basis expansion and rather rescales the whole pulse such that the shape is maintained but the amplitude is bounded by $f_{\mathrm{max}}$.

Apart from these constraints on the control pulse, one can also add in a similar way penalty terms that depend on the state of the system. E.g., one could put a penalty for populating a certain subspace or energy level. In many cases, the control pulse does not only connect the qubit levels, but can bring population also to additional levels that can be affected by dephasing or decay. If we consider the case of state transfer and call such a prohibited level $|p\rangle$, we can introduce a penalty term
\begin{eqnarray}
	J(f)= 1-F+ \lambda \int_0^T|\langle p|\psi(t)\rangle|^2dt
\end{eqnarray}
that penalizes the overall time spent in this level. It is straightforward to extend this penalty to a subspace by summing over a basis of the prohibited subspace or to introduce a similar term for gate optimization by adding a penalty term for each of the logical basis states. 

\section{Symmetries and Control Resources}

\subsection{Speed and Information Limits}
Apart from the question of controllability, i.e., the question whether a control solution exists at all, it is also important to ask whether such a solution can attain its goal within a given time and resources. In particular, the nature of the quantum mechanical equations of motion comes with a time-energy bound, the so-called quantum speed limit [\cite{Mandelstam1991,Margolus1998}]. One could also say that one cannot move from one place in the Hilbert space to another faster than given by this speed limit. For an initial state $|\psi_0\rangle$ and a target state $|\psi_G\rangle$ as well as a variance of the Hamiltonian given by $\Delta E$, this speed limit tells us that we need at least a time
\begin{eqnarray}
T\geq\frac{\arccos |\langle\psi_G|\psi_0\rangle|}{\Delta E}
\end{eqnarray}
to bring the system from the initial state to the target state. Such a speed limit can derive, e.g., from a limited control pulse amplitude or a finite interaction term for multi-qubit operations, and generalizations can be made for gate operations or open quantum systems. The quantum speed-limit can be found also by QOC [\cite{Caneva2009}]. For further reading we refer to the review by \cite{Deffner2017}.

A similar speed limit can be formulated by comparing the information content of the control pulse to the precision with which one has to steer the system in a multi-dimensional Hilbert space. In particular, one needs to encode the necessary information in the control pulse to specify the target state. To quantify this amount of information one can calculate the corresponding channel capacity. For a noiseless control with a finite resolution $\Delta f$ and maximal amplitude $f_{\mathrm{max}}$ this channel capacity is given by
Hartley's law:
\begin{equation}
	C = B \log_2 \left(1 + \frac{f_{\mathrm{max}}}{\Delta f}\right).
\end{equation}
Here, $B$ is the bandwidth of the control pulse, see \cite{Lloyd2014}.
A similar channel capacity can be found for noisy channels. The amount of information encoded in the control pulse is then limited by $CT$ and this has to match the amount of information $-D_r \log_{2}(\varepsilon)$ that is needed to distinguish any two states in the relevant Hilbert space of dimension $D_r$ up to an error $\varepsilon$. This leads to an information speed limit of
\begin{equation}\label{eq:bound_T}
	T \geq -\frac{D_r}{C}\log_{2}(\varepsilon).
\end{equation}
An alternative formulation without resorting to finite resolution of control pulse and Hilbert space can be found for considering the case of the state transfer with a parameterized control pulse. The number of the pulse parameters has to exceed the real dimension of the state space, leading to the so-called ``$2N-2$"-rule that was confirmed numerically by \cite{Moore2012,Caneva2014}.

\subsection{Entangling Gates}\label{sec:gates}

When we talk about control, especially imperfect control (i.e., we do not manage to steer the system perfectly to the desired target, but only approximately), we need a way to quantify the success of the controlled operation. We have already seen the example of state transfer, where we considered the state-transfer fidelity $F_{st}$ and the generation of a gate, where we considered the square-modulus fidelity $F_{sm}$. Here, we want to have a closer look at different ways to define a gate fidelity such that the time evolution corresponds to the target gate in various ways. We have seen that for $F_{sm}=1$ the target gate $V$ must be identical to the gate generated by the time evolution up to a difference in the global phase. If we need $U(T)\hat{=}V$ in a less strict sense, we can also consider the single-squared fidelity
\begin{eqnarray}
	F_{ss}=\frac{1}{N} \sum_{n=1}^N |\langle n | V^\dagger U(T)|n\rangle|^2\,.
\end{eqnarray}
In this case the fidelity is the sum of the elements of $V^\dagger U(T)$, each one squared separately. As a consequence, for $F_{ss}=1$, the matrix elements of  $U(T)$ and $V$ have the same absolute value but each one can have a different phase per column: $U|n\rangle=e^{i\phi_n}V|n\rangle$. This fidelity corresponds to $N$ simultaneous state transfers~[\cite{Palao2003}].
In some cases, it is desirable to fix also the global phase of the gate in the fidelity functional and thus consider the real-part fidelity
\begin{eqnarray}
F_{Re}=\frac{1}{N}\re \sum_{n=1}^N \langle n | V^\dagger U(T)|n\rangle=\frac{1}{N}\re\Tr V^\dagger U(T)\,. 
\end{eqnarray}

\paragraph{Two-Qubit Gates}
We can get more insight into the different degrees of freedom by having a closer look at two-qubit gates. We first consider two diagonal gates $U=\mathrm{diag}( \mathrm{e}^{i\phi_{U,1}}, \mathrm{e}^{i\phi_{U,2}}, \mathrm{e}^{i\phi_{U,3}}, \mathrm{e}^{i\phi_{U,4}})$ and $V=\mathrm{diag}( \mathrm{e}^{i\phi_{V,1}}, \mathrm{e}^{i\phi_{V,2}}, \mathrm{e}^{i\phi_{V,3}}, \mathrm{e}^{i\phi_{V,4}})$ and set $\Delta\phi_i := \phi_{U,i} - \phi_{V,i}$.
The gate fidelity $F_{Re}$ is then
\begin{eqnarray}
	F_{Re}=\frac{1}{4}\re\sum_{i=1}^4\mathrm{e}^{-i\Delta\phi_i}= \frac{1}{4}\sum_{i=1}^4\cos\Delta\phi_i \,\text{.}
\end{eqnarray}
We can express the phases by decomposing them into a global phase $\phi_{U,g}$, two local phases $\phi_{U,l/r}$ for the left and the right qubit and an non-local or entangling phase $\phi_{U,NL}$ via
\begin{eqnarray}
	U=\mathrm{e}^{i\left(\phi_{U,g} \id\otimes\id
		            +(\phi_{U,l}/2) \sigma_z\otimes\id
		            +(\phi_{U,r}/2)
                    \id\otimes\sigma_z
		            +(\phi_{U,NL}/4)
              \sigma_z\otimes\sigma_z
		\right)}\,.
\end{eqnarray}
We can check that $\phi_{U,NL}=\phi_{U,1}-\phi_{U,2}-\phi_{U,3}+\phi_{U,4}$ and define analogous expressions for $V$. If we give freedom to the global and local phases (e.g., by assuming that we can readjust them easily), the gate fidelity considering just the non-local contribution is
\begin{eqnarray}
	F^{NL}_{Re}=\cos\frac{\Delta\phi}{4}\label{eq:fidelity_localphase}\,,
\end{eqnarray}
where $\Delta\phi=\phi_{U,NL}-\phi_{V,NL}$. This result can be generalized to more than two qubits by considering also three- and more body operators.

A similar result can be obtained for general (non-diagonal) two-qubit gates via algebraic decomposition. In particular, $U\in SU(4)$ can be decomposed into the local operations $k_{1},k_{2}\in SU(2)\otimes SU(2)$ and the non-local part
\begin{eqnarray}
	A=\mathrm{exp}\left(\frac{i}{2}(c_{U,1}\sigma_x\otimes\sigma_x +c_{U,2}\sigma_y\otimes\sigma_y + c_{U,3} \sigma_z\otimes\sigma_z)\right)\,,
\end{eqnarray}
so that the gate can be expressed as $U=k_{1}Ak_{2}$. The non-local part is defined by the so-called Weyl coordinates $c_{U,1}$, $c_{U,2}$ and $c_{U,3}$. The same decomposition can be done for the target gate $V$. If we then assume again that we can ignore or easily realize the local transformations,  the fidelity $F_{Re}$ between $U$ and $V$ is given by
\begin{eqnarray}
	\label{eq:fidelity-unitary}F^{NL}_{Re}(U,V)&=&\max_{k_1,k_2}\frac{1}{4}\re \Tr U^\dagger k_1 V k_2 \nonumber\\
	&=&\cos\frac{\Delta c_1}{2}\cos\frac{\Delta c_2}{2}\cos\frac{\Delta c_3}{2}\,,
\end{eqnarray}
i.e., it is defined by the difference in the Weyl coordinates of $U$ and $V$, namely $\Delta c_i =c_{U,i}-c_{V,i}$ ($i=1,\dots 3$). More details on the decomposition can be found in the works of~\cite{Zhang2003a,Zhang2003b,Zhang2004a,Makhlin2002}. Its use for QOC is described by~\cite{Mueller2011,Watts2015,Goerz2015}. The advantage of this approach for QOC is linked to the arguments on information limits described in the previous section: a problem with a lower relevant dimension (e.g., the three dimensions of the Weyl chamber instead of the 15 dimensions of $SU(4)$) can be controlled with a simpler pulse and the solution is typically easier to find.

\subsection{Observables and other Control Objectives}
While state transfer and gate operations are likely the most common optimization targets, in many cases the desired target does not necessarily have to be decomposed into these basic building blocks. For example, in systems like mixed-spin systems or spin-boson systems, often only a subsystem (e.g., one of the spins) can be initialized directly into a pure state. This state could now be transferred to the other spin (subsystem), e.g., by a SWAP operation, but in the end, it is only important that the target subsystem ends up in a state as pure as possible. A natural FoM for such a task is thus the purity $\Tr \rho^2$ of the state $\rho$ of the target subsystem. Since this admits any pure state as final state of the system, it drastically reduces the control complexity.

In a similar fashion, for two systems that are only connected through a link that is hard to control (e.g., two distant nodes for quantum communication, or two units of a spatially extended quantum sensor) and the key requirement is to create entanglement, one might not want to enforce the state transfer to a specific target state (e.g., a Bell or GHZ state), but instead require to maximize only a measure of entanglement between the nodes [\cite{pezze2018}].

For quantum many-body systems, where the exponentially (in the number of constituents) size of the Hilbert space renders the direct optimization of a target state very complex, one could instead be interested only in ending up with a certain degree of entanglement (calculated by the entanglement entropy of a bipartition) or the order parameter of a phase transition (e.g., observables like the magnetization) to obtain a state that contains the desired many-body features of interest. Such FoMs for quantum many-body systems are general more straightforward to implement with gradient-free algorithms.

\subsection{Controllability and Symmetry analysis}
In Section \ref{sec:single-qubit} we have seen how we can obtain single-qubit gates from the control Hamiltonian of a single-qubit system by driving  the system alternatingly with different control phases and amplitudes. For larger systems, single-qubit gates, however, are not sufficient: for universal quantum computing, for instance, one needs to combine these single-qubit gates with two-qubit gates connecting the various qubits [\cite{barenco1995}]. Typical two-qubit gates that are often used in quantum algorithms are
\begin{eqnarray}
	\text{CNOT}=
	\begin{pmatrix}
		1 & 0 & 0 & 0\\
		0 & 1 & 0 & 0\\
		0 & 0 & 0 & 1\\
		0 & 0 & 1 & 0
	\end{pmatrix}\,,\quad
	\text{CPHASE}=
	\begin{pmatrix}
		1 & 0 & 0 & 0\\
		0 & 1 & 0 & 0\\
		0 & 0 & 1 & 0\\
		0 & 0 & 0 & \mathrm{e}^{i \phi}
	\end{pmatrix}\,,\nonumber\\
	\sqrt{\text{SWAP}}=
	\begin{pmatrix}
		1 & 0 & 0 & 0\\
		0 & \frac{\mathrm{e}^{i\frac{\pi}{4}}}{\sqrt{2}} & \frac{\mathrm{e}^{-i\frac{\pi}{4}}}{\sqrt{2}} & 0\\
		0 & \frac{\mathrm{e}^{-i\frac{\pi}{4}}}{\sqrt{2}} & \frac{\mathrm{e}^{i\frac{\pi}{4}}}{\sqrt{2}} & 0\\
		0 & 0 & 0 & 1
	\end{pmatrix}\,,\quad
	\text{SWAP}=
    \begin{pmatrix}
	1 & 0 & 0 & 0\\
	0 & 0& 1 & 0\\
	0 & 1 & 0 & 0\\
	0 & 0 & 0 & 1
   \end{pmatrix}	\,.
\end{eqnarray}
Here, CNOT, CPHASE (depending on the angle $\phi$) and $\sqrt{\text{SWAP}}$ are entangling gates that naturally arise from different types of interaction terms between two qubits or can be designed by an arbitrary interaction together with a suitable control term. The SWAP gate (realized, e.g., by two applications of $\sqrt{\text{SWAP}}$ or three applications of CNOT with alternating roles of control and target qubit) swaps the states of the two qubits but leaves the entanglement between the qubits untouched. Entangling two-qubit gates are required for example to create entanglement between two qubits, measured by entanglement measures like
\begin{eqnarray}
	\mathrm{Ent}\psi = \psi^{T} P\psi\text{, with }P=-\frac{1}{2}\sigma_{y}\otimes\sigma_{y}\,,
\end{eqnarray}
see \cite{Zhang2003a}.
We find zero entanglement, $\vert \mathrm{Ent}\psi \vert=0$, iff $\psi$ is a product state; maximum entanglement $\max_{\psi }\vert \mathrm{Ent}\psi \vert= \frac{1}{2}$ is reached, e.g., for the Bell states. We can quantify the entangling power of a gate $U$ by
\begin{eqnarray}
	\max_{\psi}\vert\mathrm{Ent}U\psi\vert \,
\end{eqnarray}
where the maximization is over all product states. The entangling power measures thus the power of a gate to turn a product state into an entangled state with corresponding value of the entanglement measure.
Considering the CPHASE gate with arbitrary angle $\phi$ and denoting it by $U(\phi)$,  e.g., we find 
\begin{eqnarray}
\max_{\psi}\vert\mathrm{Ent}U_{f}\psi\vert = \frac{1}{2}\left|\sin\frac{\phi}{2}\right|\,.
\end{eqnarray}
To see how this entangling power arises from the interaction between two qubits, we consider the simple example of a system described exclusively by an interaction of the form
\begin{eqnarray}
	H_{int}=\gamma \frac{\id-\sigma_z}{2}\otimes\frac{\id-\sigma_z}{2}\,.
\end{eqnarray}
The time evolution according to this Hamiltonian for some time $t$ leads to gates of the form
\begin{eqnarray}
	U(t)=\mathrm{diag}(1,1,1,e^{-i\gamma t})
\end{eqnarray}
and we can thus identify the entangling phase $\phi$ with the interaction strength acting over some time: $\phi=-\gamma t$.

But how do we go from here to a general two-qubit gate? With the help of Section \ref{sec:gates} we find that this one two-qubit gate in combination with arbitrary single-qubit gates is enough for generating any arbitrary two-qubit gate. Indeed, our two-qubit gate $U(t)$ has the non-local content
\begin{eqnarray}
	A_x(t)=k_1^\dagger U(t)k_2^\dagger =e^{i \frac{\gamma t}{4} \sigma_x\otimes\sigma_x}\,.
\end{eqnarray}
If we want an arbitrary non-local content, we can apply $A_x(t)$ three times with different $t$ and interleaved by local transformations that change the basis:
\begin{eqnarray}
	 A_x(t_x) k_y^\dagger A_x(t_y) k_y k_z^\dagger A_x(t_z)k_z=e^{i \frac{\gamma t_x}{4} \sigma_x\otimes\sigma_x + i \frac{\gamma t_y}{4} \sigma_y\otimes\sigma_y + i \frac{\gamma t_z}{4} \sigma_z\otimes\sigma_z}
\end{eqnarray}
with the local transformations $k_z=e^{-i\frac{\pi}{4}\sigma_y}\otimes e^{-i\frac{\pi}{4}\sigma_y}$ and $k_y=e^{i\frac{\pi}{4}\sigma_z}\otimes e^{i\frac{\pi}{4}\sigma_z}$.

Summarizing, we can see that we can obtain arbitrary gates by interleaving gates that are directly achievable with the given resources. One can more formally investigate this by considering unitary transformations as elements of a Lie group and by considering the terms in the Hamiltonian as elements of the corresponding Lie algebra or generators [\cite{jurdjevic1972}]. A given term, i.e., $H_{int}$ in our example, can move the system in a certain direction of the Lie algebra. By interleaving the interaction with different (in the example local) gates, one can effectively change this direction in the Lie algebra. In our example we have written down a small sequence of gates corresponding to "straight lines" of finite length in the Lie group, leading to a discretized path. In principle, one can also continuously change the direction in the Lie group by slowly modulating the control term of a Hamiltonian
\begin{eqnarray}
	H=H_0+f(t)H_c\,.
\end{eqnarray}
For every choice of $f(t)$ we create a different element of the Lie algebra -- out of the two-dimensional subspace spanned by $H_0$ and $H_c$. However, the total time evolution is not limited to this subspace, because moving along two such directions $H_1$ and $H_2$ will result in a third one given by the commutator $[H_1,H_2]=H_1H_2-H_2H_1$. According to the Baker-Campbell-Hausdorff formula (for small steps in the respective directions), we obtain
\begin{eqnarray}
	e^{-i H_1 t_1} e^{-iH_2 t_2}\approx e^{-i H_1 t_1 - i H_2 t_2 - \frac{1}{2}[H_1,H_2]t_1t_2}\,.
\end{eqnarray}
Note that $[H_1,H_2]$ is not necessarily an element of the subspace spanned by $H_1$ and $H_2$. This allows us to enlarge the accessible dimensions of the Lie algebra. Let's go back to our example. If we consider $H_0=H_{int}=\gamma \frac{\id-\sigma_z}{2}\otimes\frac{\id-\sigma_z}{2}$ and $H_c=\sigma_x\otimes\id$ (i.e., the possibility to drive qubit 1), we can directly activate the Lie algebra elements
$$\mathfrak{a}_1=-i\frac{\id-\sigma_z}{2}\otimes\frac{\id-\sigma_z}{2}\quad\text{and}\quad \mathfrak{a}_2=-i\sigma_x\otimes\id$$
The commutator is
$$\mathfrak{a}_3=\left[\mathfrak{a}_1,\mathfrak{a}_2 \right]
= i\sigma_y\otimes\frac{\id-\sigma_z}{2}\,.$$
The next order of commutators are
$$\mathfrak{a}_4=\left[\mathfrak{a}_1,\mathfrak{a}_3 \right]
= i\sigma_x\otimes \frac{\id-\sigma_z}{2}$$
and
$$\mathfrak{a}_5=\left[\mathfrak{a}_2,\mathfrak{a}_3 \right]
= 2i\sigma_z\otimes \frac{\id-\sigma_z}{2}\,.$$
From here, we have $[\mathfrak{a}_1,\mathfrak{a}_4 ]=-\mathfrak{a}_3$, $[\mathfrak{a}_1,\mathfrak{a}_5 ]=0$, $[\mathfrak{a}_2,\mathfrak{a}_4 ]=0$, $[\mathfrak{a}_2,\mathfrak{a}_5 ]=4\mathfrak{a}_3$, $[\mathfrak{a}_3,\mathfrak{a}_4 ]=\mathfrak{a}_5$, $[\mathfrak{a}_3,\mathfrak{a}_5 ]=4\mathfrak{a}_4$ and finally $[\mathfrak{a}_4,\mathfrak{a}_5 ]=4\mathfrak{a}_3$ and thus do not find any new terms. The available terms in the Hamiltonian thus allow us to access five independent directions in the Lie algebra. One can show with an analogous calculation that instead, if we add a second control term $f_2(t)\id\otimes \sigma_x$, one can obtain access to the full Lie algebra of $\mathfrak{su}(4)$ and thus generate any arbitrary two-qubit gate.

Indeed, one can formulate a general theorem for controllability of a quantum system. Let us consider a $d$-dimensional Hilbert space and the Hamiltonian
\begin{eqnarray}
	H=H_0+f(t)H_c\,.
\end{eqnarray}
The system is called \emph{controllable} if for every unitary $U\in SU(d)$ there is a finite time $T$ and a control pulse $f(t)$ such that $U=U_f(T)$, where $U_f(T)$ is the unitary resulting from the time evolution of the system according to $f(t)$. The theorem now states that a system is controllable iff the dynamical Lie algebra (i.e., the operators $H_0$, $H_c$ and their nested commutators) is of full rank, i.e., spans $\mathfrak{su}(d)$,  see~\cite{altafini2002, DAlessandro2007}. Instead of (full) controllability, one can also consider weaker concepts such as the question of reachability of a given gate or state or a subset of gates/states. In many cases, even approximate reachability (i.e., the question whether one can achieve the control target up to a certain tolerable error) is sufficient for reaching the target with the desired fidelity.

\section{Conclusions and Further Reading}
In this chapter we have introduced the reader to the topic of Quantum Control by first looking at simple examples that at the same time present the main challenges appearing in many more complex applications and give an intuitive understanding of the control strategies. Controlled quantum operations can be subject to errors coming from system or model imperfections, from pulse imperfections or from the coupling to degrees of freedom that are not controlled, and controlled quantum operations on a subsystem can introduce errors on other parts of a larger system such as cross-talk or phase shifts. In more complex systems the simple control strategies presented in the earlier sections might have to be combined or supplemented by heavier machinery. We have discussed methods of adiabatic control such as shortcuts to adiabaticity [see \cite{Guery-Odelin2019}] and DRAG [see \cite{Theis2018}]. We have further discussed composite pulses and pulse sequences or dynamical decoupling, that can often be applied when time scales between control and drift are separated such as in nuclear spin systems, see \cite{Vandersypen2005,Viola1999,Suter2016}.
We then moved to numerical algorithms for QOC and machine learning. Here, not only many reviews can be recommended, such as the ones by \cite{Glaser2015} and \cite{Koch2022} that show a pathway to quantum technology, the ones by \cite{Brif2010}, \cite{Rembold2020}, \cite{Koch2025}, that give a more practical introduction, or the ones by \cite{Koch2016} or \cite{Mueller2022} that focus on open quantum system or a particular algorithm (CRAB), respectively. There are also numerical toolboxes available, such as QuTiP by \cite{Johansson2013}, Krotov by \cite{Goerz2019} or QuOCS by \cite{Rossignolo2023}.

Finally, we discussed ways to cast control objectives efficiently into FoMs, e.g., for gate optimization: we can refer the reader to the reviews by \cite{Mueller2022} and by \cite{Koch2025} for further reading. For a more rigorous treatment of controllability of unitary systems we refer to the book of \cite{DAlessandro2007} and for open systems again to the review by \cite{Koch2016}.

There are also many topics that are part of quantum control but that we could not discuss here. While we have discussed closed-loop control, where feedback from measurements at final time is used to iteratively optimize a control pulse, one could also use real-time feedback to control a quantum system, see \cite{Rouchon2022}. Furthermore, for QOC we have discussed just numerical algorithms, while in many cases also analytic solutions to the optimization problem can be found via the Pontryagin maximum principle, see \cite{Boscain2021}.

Concluding, we hope that this chapter can help newcomers to quantum control to get a flavor of the challenges and methods, and help them to learn to apply them to real-world challenges. For the practitioner, we hope that our work gives a good summary of how broad the field of quantum control has become and it might as well serve as a starting point for further dissemination of the knowledge to the emerging profession of quantum engineers.

\section*{Acknowledgements}
We acknowledge valuable discussions with Robert Zeier. Part of the material has been developed for our lectures at the University of Cologne.

\bibliographystyle{elsarticle-harv}

\bibliography{Refs}

@article{Leibfried2003,
  title = {Quantum dynamics of single trapped ions},
  author = {Leibfried, D. and Blatt, R. and Monroe, C. and Wineland, D.},
  journal = {Rev. Mod. Phys.},
  volume = {75},
  issue = {1},
  pages = {281--324},
  numpages = {0},
  year = {2003},
  month = {Mar},
  publisher = {American Physical Society},
  doi = {10.1103/RevModPhys.75.281}
}

@article{Doria2011,
	Author = {Doria, Patrick and Calarco, Tommaso and Montangero, Simone},
	Doi = {10.1103/PhysRevLett.106.190501},
	Issue = {19},
	Journal = {Phys. Rev. Lett.},
	Month = {May},
	Numpages = {4},
	Pages = {190501},
	Publisher = {American Physical Society},
	Title = {Optimal Control Technique for Many-Body Quantum Dynamics},
	Volume = {106},
	Year = {2011}
}

@incollection{Mandelstam1991,
author={Mandelstam, L. and Tamm, Ig.},
title={The Uncertainty Relation Between Energy and Time in Non-relativistic Quantum Mechanics},
bookTitle={Selected Papers},
year={1991},
publisher={Springer Berlin Heidelberg},
address={Berlin, Heidelberg},
pages={115--123},
isbn={978-3-642-74626-0},
doi= {10.1007/978-3-642-74626-0\_8}
}

@article{Schrieffer1966,
  author = {Schrieffer, J. R. and Wolff, P. A.},
  title = {Relation between the Anderson and Kondo Hamiltonians},
  journal = {Physical Review},
  volume = {149},
  number = {2},
  pages = {491--492},
  year = {1966},
  doi = {10.1103/PhysRev.149.491}
}

@article{Deffner2017,
  author = {Sebastian Deffner and Steve Campbell},
  title = {Quantum speed limits: from Heisenberg's uncertainty principle to optimal quantum control},
  journal = {Journal of Physics A: Mathematical and Theoretical},
  volume = {50},
  number = {45},
  pages = {453001},
  year = {2017},
  month = {oct},
  publisher = {IOP Publishing},
  doi = {10.1088/1751-8121/aa86c6}
}

@article{Caneva2009,
  author = {Caneva, T. and Murphy, M. and Calarco, T. and Fazio, R. and Montangero, S. and Giovannetti, V. and Santoro, G. E.},
  title = {Optimal Control at the Quantum Speed Limit},
  journal = {Phys. Rev. Lett.},
  volume = {103},
  number = {24},
  pages = {240501},
  year = {2009},
  month = {Dec},
  publisher = {American Physical Society},
  doi = {10.1103/PhysRevLett.103.240501}
}

@article{Margolus1998,
  author = {Norman Margolus and Lev B. Levitin},
  title = {The maximum speed of dynamical evolution},
  journal = {Physica D: Nonlinear Phenomena},
  volume = {120},
  number = {1},
  pages = {188--195},
  year = {1998},
  doi = {10.1016/S0167-2789(98)00054-2}
}

@article{Pagano2024,
  author = {Pagano, Alice and M{\"u}ller, Matthias M. and Calarco, Tommaso and Montangero, Simone and Rembold, Phila},
  title = {Role of bases in quantum optimal control},
  journal = {Phys. Rev. A},
  volume = {110},
  number = {6},
  pages = {062608},
  year = {2024},
  month = {Dec},
  publisher = {American Physical Society},
  doi = {10.1103/PhysRevA.110.062608}
}

@article{Caneva2011,
  author = {Caneva, Tommaso and Calarco, Tommaso and Montangero, Simone},
  title = {Chopped random-basis quantum optimization},
  journal = {Phys. Rev. A},
  volume = {84},
  number = {2},
  pages = {022326},
  year = {2011},
  month = {Aug},
  publisher = {American Physical Society},
  doi = {10.1103/PhysRevA.84.022326}
}

@article{Hahn1950,
  author = {Hahn, E. L.},
  title = {Spin Echoes},
  journal = {Phys. Rev.},
  volume = {80},
  number = {4},
  pages = {580--594},
  year = {1950},
  month = {Nov},
  publisher = {American Physical Society},
  doi = {10.1103/PhysRev.80.580}
}

@article{Waugh1968,
  author = {Waugh, J. S. and Huber, L. M. and Haeberlen, U.},
  title = {Approach to High-Resolution NMR in Solids},
  journal = {Phys. Rev. Lett.},
  volume = {20},
  number = {5},
  pages = {180--182},
  year = {1968},
  month = {Jan},
  publisher = {American Physical Society},
  doi = {10.1103/PhysRevLett.20.180}
}

@article{Defouquieres2011,
  author = {P. de Fouquieres and S.G. Schirmer and S.J. Glaser and Ilya Kuprov},
  title = {Second order gradient ascent pulse engineering},
  journal = {Journal of Magnetic Resonance},
  volume = {212},
  number = {2},
  pages = {412--417},
  year = {2011},
  doi = {10.1016/j.jmr.2011.07.023}
}

@article{Soerensen2018,
  author = {S{\o}rensen, J. J. W. H. and Aranburu, M. O. and Heinzel, T. and Sherson, J. F.},
  title = {Quantum optimal control in a chopped basis: Applications in control of Bose-Einstein condensates},
  journal = {Phys. Rev. A},
  volume = {98},
  number = {2},
  pages = {022119},
  year = {2018},
  month = {Aug},
  publisher = {American Physical Society},
  doi = {10.1103/PhysRevA.98.022119}
}

@article{Casanova2015,
  author = {Casanova, J. and Wang, Z.-Y. and Haase, J. F. and Plenio, M. B.},
  title = {Robust dynamical decoupling sequences for individual-nuclear-spin addressing},
  journal = {Phys. Rev. A},
  volume = {92},
  number = {4},
  pages = {042304},
  year = {2015},
  month = {Oct},
  publisher = {American Physical Society},
  doi = {10.1103/PhysRevA.92.042304}
}

@article{Rossignolo2023,
  author = {Marco Rossignolo and Thomas Reisser and Alastair Marshall and Phila Rembold and Alice Pagano and Philipp J. Vetter and Ressa S. Said and Matthias M. M{\"u}ller and Felix Motzoi and Tommaso Calarco and Fedor Jelezko and Simone Montangero},
  title = {QuOCS: The quantum optimal control suite},
  journal = {Computer Physics Communications},
  volume = {291},
  pages = {108782},
  year = {2023},
  doi = {10.1016/j.cpc.2023.108782}
}

@article{Khaneja2005,
  author = {Navin Khaneja and Timo Reiss and Cindie Kehlet and Thomas Schulte-Herbr{\"u}ggen and Steffen J. Glaser},
  title = {Optimal control of coupled spin dynamics: design of NMR pulse sequences by gradient ascent algorithms},
  journal = {Journal of Magnetic Resonance},
  volume = {172},
  number = {2},
  pages = {296--305},
  year = {2005},
  doi = {10.1016/j.jmr.2004.11.004}
}

@article{Boscain2021,
  title = {Introduction to the Pontryagin Maximum Principle for Quantum Optimal Control},
  author = {Boscain, U. and Sigalotti, M. and Sugny, D.},
  journal = {PRX Quantum},
  volume = {2},
  issue = {3},
  pages = {030203},
  numpages = {31},
  year = {2021},
  month = {Sep},
  publisher = {American Physical Society},
  doi = {10.1103/PRXQuantum.2.030203}
}

@article{Brif2010,
  Author = {Constantin Brif and Raj Chakrabarti and Herschel Rabitz},
  Doi = {10.1088/1367-2630/12/7/075008},
  Journal = {New Journal of Physics},
  Month = {jul},
  Number = {7},
  Pages = {075008},
  Publisher = {{IOP} Publishing},
  Title = {Control of quantum phenomena: past, present and future},
  Volume = {12},
  Year = 2010
}

@article{Choi2020,
  title = {Robust Dynamic Hamiltonian Engineering of Many-Body Spin Systems},
  author = {Choi, Joonhee and Zhou, Hengyun and Knowles, Helena S. and Landig, Renate and Choi, Soonwon and Lukin, Mikhail D.},
  journal = {Phys. Rev. X},
  volume = {10},
  issue = {3},
  pages = {031002},
  numpages = {27},
  year = {2020},
  month = {Jul},
  publisher = {American Physical Society},
  doi = {10.1103/PhysRevX.10.031002}
}

@article{georgiadis2026,
  title={Simple analytical flux-tuned iSWAP pulses for leakage suppression},
  author={Georgiadis, Dimitrios and Li, Boxi and Galicia, Asier and Barends, Rami and C{\'a}rdenas-L{\'o}pez, FA and Motzoi, Felix},
  journal={arXiv preprint arXiv:2606.13052},
  year={2026},
  doi = {10.48550/arXiv.2606.13052}
}

@article{Johansson2013,
  title = {QuTiP 2: A Python framework for the dynamics of open quantum systems},
  journal = {Computer Physics Communications},
  volume = {184},
  number = {4},
  pages = {1234-1240},
  year = {2013},
  issn = {0010-4655},
  doi = {10.1016/j.cpc.2012.11.019},
  author = {J.R. Johansson and P.D. Nation and Franco Nori}
}

@article{Kirchhoff2025,
  title = {Correction Formulas for the M\o{}lmer-S\o{}rensen Gate Under Strong Driving},
  author = {Kirchhoff, Susanna and Wilhelm, Frank K. and Motzoi, Felix},
  journal = {PRX Quantum},
  volume = {6},
  issue = {1},
  pages = {010328},
  numpages = {16},
  year = {2025},
  month = {Feb},
  publisher = {American Physical Society},
  doi = {10.1103/PRXQuantum.6.010328}
}

@article{li2025universal,
  title={Universal pulses for superconducting qudit ladder gates},
  author={Li, Boxi and C{\'a}rdenas-L{\'o}pez, FA and Lupascu, Adrian and Motzoi, Felix},
  journal={PRX Quantum},
  volume={6},
  number={3},
  pages={030357},
  year={2025},
  publisher={APS},
  doi = {10.1103/9dxw-4c7y}
}

@article{Lloyd2014,
  Author = {Lloyd, S. and Montangero, S.},
  Issue = {1},
  Journal = {Phys. Rev. Lett.},
  Month = {Jul},
  Numpages = {5},
  Pages = {010502},
  Publisher = {American Physical Society},
  Title = {Information Theoretical Analysis of Quantum Optimal Control},
  Volume = {113},
  Year = {2014},
  Doi = {10.1103/PhysRevLett.113.010502}
}

@article{Mueller2022,
  doi = {10.1088/1361-6633/ac723c},
  year = {2022},
  month = {jun},
  publisher = {IOP Publishing},
  volume = {85},
  number = {7},
  pages = {076001},
  author = {Matthias M Müller and Ressa S Said and Fedor Jelezko and Tommaso Calarco and Simone Montangero},
  title = {One decade of quantum optimal control in the chopped random basis},
  journal = {Reports on Progress in Physics}
}

@article{Reich2012,
  title={Monotonically convergent optimization in quantum control using Krotov's method},
  volume = {136},
  issn = {0021-9606},
  doi = {10.1063/1.3691827},
  number = {10},
  journal = {The Journal of Chemical Physics},
  author = {Reich, Daniel M. and Ndong, Mamadou and Koch, Christiane P.},
  month = mar,
  year = {2012},
  pages = {104103}
}

@article{Rouchon2022,
  title = {A tutorial introduction to quantum stochastic master equations based on the qubit/photon system},
  journal = {Annual Reviews in Control},
  volume = {54},
  pages = {252-261},
  year = {2022},
  issn = {1367-5788},
  doi = {10.1016/j.arcontrol.2022.09.006},
  author = {Pierre Rouchon}
}

@article{Soerensen2000,
  title = {Entanglement and quantum computation with ions in thermal motion},
  author = {S\o{}rensen, Anders and M\o{}lmer, Klaus},
  journal = {Phys. Rev. A},
  volume = {62},
  issue = {2},
  pages = {022311},
  numpages = {11},
  year = {2000},
  month = {Jul},
  publisher = {American Physical Society},
  doi = {10.1103/PhysRevA.62.022311}
}

@article{Suter2016,
  title = {Colloquium: Protecting quantum information against environmental noise},
  author = {Suter, Dieter and \'Alvarez, Gonzalo A.},
  journal = {Rev. Mod. Phys.},
  volume = {88},
  issue = {4},
  pages = {041001},
  numpages = {23},
  year = {2016},
  month = {Oct},
  publisher = {American Physical Society},
  doi = {10.1103/RevModPhys.88.041001}
}

@article{Theis2018,
  doi = {10.1209/0295-5075/123/60001},
  year = {2018},
  month = {oct},
  publisher = {EDP Sciences, IOP Publishing and Società Italiana di Fisica},
  volume = {123},
  number = {6},
  pages = {60001},
  author = {Theis, L. S. and Motzoi, F. and Machnes, S. and Wilhelm, F. K.},
  title = {Counteracting systems of diabaticities using DRAG controls: The status after 10 years(a)},
  journal = {Europhysics Lett.}
}

@article{Viola1999,
  title = {Dynamical Decoupling of Open Quantum Systems},
  author = {Viola, Lorenza and Knill, Emanuel and Lloyd, Seth},
  journal = {Phys. Rev. Lett.},
  volume = {82},
  issue = {12},
  pages = {2417--2421},
  year = {1999},
  month = {Mar},
  publisher = {American Physical Society},
  doi = {10.1103/PhysRevLett.82.2417}
}

@article{Glaser2015,
  Author = {Glaser, Steffen J. and Boscain, Ugo and Calarco, Tommaso and Koch, Christiane P. and K{\"o}ckenberger, Walter and Kosloff, Ronnie and Kuprov, Ilya and Luy, Burkhard and Schirmer, Sophie and Schulte-Herbr{\"u}ggen, Thomas and Sugny, Dominique and Wilhelm, Frank K.},
  Day = {17},
  Doi = {10.1140/epjd/e2015-60464-1},
  Issn = {1434-6079},
  Journal = {The European Physical Journal D},
  Month = {Dec},
  Number = {12},
  Pages = {279},
  Title = {Training Schr{\"o}dinger's cat: quantum optimal control},
  Volume = {69},
  Year = {2015}
}

@article{Guery-Odelin2019,
  title = {Shortcuts to adiabaticity: Concepts, methods, and applications},
  author = {Gu\'ery-Odelin, D. and Ruschhaupt, A. and Kiely, A. and Torrontegui, E. and Mart\'{\i}nez-Garaot, S. and Muga, J. G.},
  journal = {Rev. Mod. Phys.},
  volume = {91},
  issue = {4},
  pages = {045001},
  numpages = {54},
  year = {2019},
  month = {Oct},
  publisher = {American Physical Society},
  doi = {10.1103/RevModPhys.91.045001}
}

@article{Koch2016,
  Author = {Christiane P Koch},
  Doi = {10.1088/0953-8984/28/21/213001},
  Journal = {Journal of Physics: Condensed Matter},
  Month = {may},
  Number = {21},
  Pages = {213001},
  Publisher = {{IOP} Publishing},
  Title = {Controlling open quantum systems: tools, achievements, and limitations},
  Volume = {28},
  Year = {2016}
}

@article{Koch2022,
  title = {Quantum optimal control in quantum technologies. Strategic report on current status, visions and goals for research in Europe},
  volume = {9},
  issn = {2196-0763},
  doi = {10.1140/epjqt/s40507-022-00138-x},
  number = {1},
  journal = {EPJ Quantum Technology},
  author = {Koch, Christiane P. and Boscain, Ugo and Calarco, Tommaso and Dirr, Gunther and Filipp, Stefan and Glaser, Steffen J. and Kosloff, Ronnie and Montangero, Simone and Schulte-Herbrüggen, Thomas and Sugny, Dominique and Wilhelm, Frank K.},
  month = jul,
  year = {2022},
  pages = {19}
}

@article{Koch2025,
  title = {Introduction to quantum control: From basic concepts to applications in quantum technologies},
  author = {Christiane P. Koch},
  year = {2025},
  eprint = {2512.04990},
  archivePrefix = {arXiv},
  primaryClass = {quant-ph},
 doi={10.48550/arXiv.2512.04990},
  journal = {arXiv preprint}
}

@article{Konnov1999,
  Author = {Konnov, A.~I. and Krotov, V.~F.},
  Issue = {10},
  Journal = {Autom. Remote Control},
  Pages = {1427--1436},
  Title = {On global methods for the successive improvement of control processes},
  Volume = {60},
  Year = {1999}
}

@article{Makhlin2002,
  Author = {Makhlin, Yuriy},
  Da = {2002/08/01},
  Id = {Makhlin2002},
  Isbn = {1573-1332},
  Journal = {Quantum Information Processing},
  Number = {4},
  Pages = {243--252},
  Title = {Nonlocal Properties of Two-Qubit Gates and Mixed States, and the Optimization of Quantum Computations},
  Volume = {1},
  Year = {2002},
 doi={10.1023/A\%3A1022144002391}
}

@article{Motzoi2009,
  title = {Simple Pulses for Elimination of Leakage in Weakly Nonlinear Qubits},
  author = {Motzoi, F. and Gambetta, J. M. and Rebentrost, P. and Wilhelm, F. K.},
  journal = {Phys. Rev. Lett.},
  volume = {103},
  issue = {11},
  pages = {110501},
  numpages = {4},
  year = {2009},
  month = {Sep},
  publisher = {American Physical Society},
  doi = {10.1103/PhysRevLett.103.110501}
}

@article{Palao2003,
  Author = {Palao, Jos\'e P. and Kosloff, Ronnie},
  Doi = {10.1103/PhysRevA.68.062308},
  Issue = {6},
  Journal = {Phys. Rev. A},
  Month = {Dec},
  Numpages = {13},
  Pages = {062308},
  Publisher = {American Physical Society},
  Title = {Optimal control theory for unitary transformations},
  Volume = {68},
  Year = {2003}
}

@article{Palao2008,
  Author = {Palao, Jos\'e P. and Kosloff, Ronnie and Koch, Christiane P.},
  Doi = {10.1103/PhysRevA.77.063412},
  Issue = {6},
  Journal = {Phys. Rev. A},
  Month = {Jun},
  Numpages = {11},
  Pages = {063412},
  Publisher = {American Physical Society},
  Title = {Protecting coherence in optimal control theory: State-dependent constraint approach},
  Volume = {77},
  Year = {2008}
}

@article{Rembold2020,
  author = {Rembold, Phila and Oshnik, Nimba and Müller, Matthias M. and Montangero, Simone and Calarco, Tommaso and Neu, Elke},
  title = {Introduction to quantum optimal control for quantum sensing with nitrogen-vacancy centers in diamond},
  journal = {AVS Quantum Science},
  volume = {2},
  number = {2},
  pages = {024701},
  year = {2020},
  doi = {10.1116/5.0006785}
}

@article{Sklarz2002,
  Author = {Sklarz, Shlomo E. and Tannor, David J.},
  Doi = {10.1103/PhysRevA.66.053619},
  Issue = {5},
  Journal = {Phys. Rev. A},
  Month = {Nov},
  Numpages = {9},
  Pages = {053619},
  Publisher = {American Physical Society},
  Title = {Loading a Bose-Einstein condensate onto an optical lattice: An application of optimal control theory to the nonlinear Schr\"odinger equation},
  Volume = {66},
  Year = {2002}
}

@article{Vandersypen2005,
  title = {NMR techniques for quantum control and computation},
  author = {Vandersypen, L. M. K. and Chuang, I. L.},
  journal = {Rev. Mod. Phys.},
  volume = {76},
  issue = {4},
  pages = {1037--1069},
  year = {2005},
  month = {Jan},
  publisher = {American Physical Society},
  doi = {10.1103/RevModPhys.76.1037}
}

@article{Wimperis1994,
  title = {Broadband, Narrowband, and Passband Composite Pulses for Use in Advanced NMR Experiments},
  journal = {Journal of Magnetic Resonance, Series A},
  volume = {109},
  number = {2},
  pages = {221-231},
  year = {1994},
  issn = {1064-1858},
  doi = {10.1006/jmra.1994.1159},
  author = {S. Wimperis}
}

@article{an2021,
  title={Quantum optimal control of multilevel dissipative quantum systems with reinforcement learning},
  author={An, Zheng and Song, Hai-Jing and He, Qi-Kai and Zhou, DL},
  journal={Phys. Rev. A},
  volume={103},
  number={1},
  pages={012404},
  year={2021},
  publisher={APS},
 DOI= {10.1103/PhysRevA.103.012404}
}

@article{bukov2018,
  title={Reinforcement learning for autonomous preparation of Floquet-engineered states: Inverting the quantum Kapitza oscillator},
  author={Bukov, Marin},
  journal={Phys. Rev. B},
  volume={98},
  number={22},
  pages={224305},
  year={2018},
  publisher={APS},
  doi= {10.1103/PhysRevB.98.224305}
}

@article{calzavara2025,
  title={Classical feature map surrogates and metrics for quantum control landscapes},
  author={Calzavara, Martino and Calarco, Tommaso and Motzoi, Felix},
  journal={arXiv:2509.25930},
  year={2025},
  doi={10.48550/arXiv.2509.25930}
}

@article{Caneva2014,
  title={Complexity of controlling quantum many-body dynamics},
  author={Caneva, T. and Silva, A. and Fazio, R. and Lloyd, S. and Calarco, T. and Montangero, S.},
  journal={Phys. Rev. A},
  volume={89},
  number={4},
  pages={042322},
  year={2014},
  publisher={American Physical Society},
  doi={10.1103/PhysRevA.89.042322}
}

@article{Cummins2000,
  title={Use of composite rotations to correct systematic errors in NMR quantum computation},
  author={H K Cummins and J A Jones},
  journal={New Journal of Physics},
  volume={2},
  number={1},
  pages={6},
  year={2000},
  publisher={IOP Publishing},
  doi={10.1088/1367-2630/2/1/006}
}

@article{Cummins2003,
  title={Tackling systematic errors in quantum logic gates with composite rotations},
  author={Cummins, Holly K. and Llewellyn, Gavin and Jones, Jonathan A.},
  journal={Phys. Rev. A},
  volume={67},
  number={4},
  pages={042308},
  year={2003},
  publisher={APS},
  doi={10.1103/PhysRevA.67.042308}
}

@book{DAlessandro2007,
  title={Introduction to Quantum Control and Dynamics},
  author={D'Alessandro, Domenico},
  publisher={Chapman \& Hall},
  isbn={9781584888840},
  pages={1--360},
  year={2007},
  doi={10.1201/9781003051268-1}
}

@article{dalgaard2020,
  title={Hessian-based optimization of constrained quantum control},
  author={Dalgaard, Mogens and Motzoi, Felix and Jensen, Jesper Hasseriis Mohr and Sherson, Jacob},
  journal={Phys. Rev. A},
  volume={102},
  number={4},
  pages={042612},
  year={2020},
  publisher={APS},
doi = {10.1103/PhysRevA.102.042612}
}

@article{dalgaard2020b,
  title={Global optimization of quantum dynamics with AlphaZero deep exploration},
  author={Dalgaard, Mogens and Motzoi, Felix and S{\o}rensen, Jens Jakob and Sherson, Jacob},
  journal={NPJ Quantum Information},
  volume={6},
  number={1},
  pages={6},
  year={2020},
  publisher={Nature Publishing Group UK London},
  doi={10.1038/s41534-019-0241-0}
}

@article{dalgaard2022,
  title={Predicting quantum dynamical cost landscapes with deep learning},
  author={Dalgaard, Mogens and Motzoi, Felix and Sherson, Jacob},
  journal={Phys. Rev. A},
  volume={105},
  number={1},
  pages={012402},
  year={2022},
  publisher={APS},
  doi={10.1103/PhysRevA.105.012402}
}

@article{del2013,
  title={Shortcuts to adiabaticity by counterdiabatic driving},
  author={Del Campo, Adolfo},
  journal={Phys. Rev. Lett.},
  volume={111},
  number={10},
  pages={100502},
  year={2013},
  publisher={APS},
  doi={10.1103/PhysRevLett.111.100502
Focus to learn more}
}

@article{demirplak2003,
  title={Adiabatic population transfer with control fields},
  author={Demirplak, Mustafa and Rice, Stuart A},
  journal={The Journal of Physical Chemistry A},
  volume={107},
  number={46},
  pages={9937--9945},
  year={2003},
  publisher={ACS Publications},
  doi={10.1021/jp030708a}
}

@article{ernst2025,
  title={Reinforcement learning for quantum control under physical constraints},
  author={Ernst, Jan Ole and Chatterjee, Aniket and Franzmeyer, Tim and Kuhn, Axel},
  journal={arXiv:2501.14372},
  year={2025},
  doi={10.48550/arXiv.2501.14372}
}

@article{Goerz2015,
  title={Optimizing for an arbitrary perfect entangler. II. Application},
  author={Goerz, Michael H. and Gualdi, Giulia and Reich, Daniel M. and Koch, Christiane P. and Motzoi, Felix and Whaley, K. Birgitta and Vala, Jiri and M{\"u}ller, Matthias M. and Montangero, Simone and Calarco, Tommaso},
  journal={Phys. Rev. A},
  volume={91},
  number={6},
  pages={062307},
  year={2015},
  publisher={APS},
  doi={10.1103/PhysRevA.91.062307}
}

@article{Goerz2019,
  title={Krotov: A Python implementation of Krotov's method for quantum optimal control},
  author={Michael H. Goerz and Daniel Basilewitsch and Fernando Gago-Encinas and Matthias G. Krauss and Karl P. Horn and Daniel M. Reich and Christiane P. Koch},
  journal={SciPost Physics},
  volume={7},
  pages={080},
  year={2019},
  doi={10.21468/SciPostPhys.7.6.080}
}

@article{Jarzynski2013,
  title={Generating shortcuts to adiabaticity in quantum and classical dynamics},
  author={Jarzynski, Christopher},
  journal={Phys. Rev. A},
  volume={88},
  number={4},
  pages={040101(R)},
  year={2013},
  publisher={APS},
  doi={10.1103/PhysRevA.88.040101}
}

@article{jerger2024,
  title = {Dispersive Qubit Readout with Intrinsic Resonator Reset},
  author = {Jerger, M. and Motzoi, F. and Gao, Y. and Dickel, C. and Buchmann, L. and C{\'a}rdenas-L{\'o}pez, F. A. and Bengtsson, A. and Tancredi, G. and Warren, C. W. and Bylander, J. and DiVincenzo, D. and Barends, R. and Bushev, P. A.},
  journal = {PRX Quantum},
  volume = {7},
  issue = {3},
  pages = {033042},
  numpages = {14},
  year = {2026},
  month = {Aug},
  publisher = {American Physical Society},
  doi = {10.1103/z5x1-2mr9}
}

@article{jesus2025,
  title = {Analytical Blueprint for 99.999\% Fidelity X-Gates on Present Superconducting Hardware Under Strong Driving},
  author = {Jesus, Jos{\'e} Diogo Da Costa and Li, Boxi and Gao, Yuan and Barends, Rami and C{\'a}rdenas-L{\'o}pez, Francisco Andr{\'e}s and Motzoi, Felix},
  journal = {PRX Quantum},
  volume = {7},
  issue = {3},
  pages = {033040},
  numpages = {17},
  year = {2026},
  month = {Aug},
  publisher = {American Physical Society},
  doi = {10.1103/sdxb-v39h}
}

@article{Koch2007Transmon,
  author  = {Koch, Jens and Yu, Terri M. and Gambetta, Jay
             and Houck, A. A. and Schuster, D. I.
             and Majer, J. and Blais, Alexandre
             and Devoret, M. H. and Girvin, S. M.
             and Schoelkopf, R. J.},
  title   = {Charge-insensitive qubit design derived
             from the Cooper pair box},
  journal = {Physical Review A},
  volume  = {76},
  number  = {4},
  pages   = {042319},
  year    = {2007},
  doi     = {10.1103/PhysRevA.76.042319}
}

@article{kuklinski1989,
  title = {Adiabatic population transfer in a three-level system driven by delayed laser pulses},
  author = {Kuklinski, J. R. and Gaubatz, U. and Hioe, F. T. and Bergmann, K.},
  journal = {Phys. Rev. A},
  volume = {40},
  issue = {11},
  pages = {6741(R)--6744(R)},
  numpages = {0},
  year = {1989},
  month = {Dec},
  publisher = {American Physical Society},
  doi = {10.1103/PhysRevA.40.6741}
}

@article{Moore2012,
  title={Exploring constrained quantum control landscapes},
  author={Moore, Katharine W. and Rabitz, Herschel},
  journal={The Journal of Chemical Physics},
  volume={137},
  number={13},
  pages={134113},
  year={2012},
  doi= {10.1063/1.4757133}
}

@article{Mueller2011,
  title={Optimizing entangling quantum gates for physical systems},
  author={M{\"u}ller, M. M. and Reich, D. M. and Murphy, M. and Yuan, H. and Vala, J. and Whaley, K. B. and Calarco, T. and Koch, C. P.},
  journal={Phys. Rev. A},
  volume={84},
  number={4},
  pages={042315},
  year={2011},
  doi= {10.1103/PhysRevA.84.042315}
}

@article{motzoi2011,
  title={Optimal control methods for rapidly time-varying Hamiltonians},
  author={Motzoi, F and Gambetta, Jay M and Merkel, Seth T and Wilhelm, Frank K},
  journal={Phys. Rev. A},
  volume={84},
  number={2},
  pages={022307},
  year={2011},
  publisher={APS},
  doi={10.1103/PhysRevA.84.022307}
}

@article{motzoi2013,
  title={Improving frequency selection of driven pulses using derivative-based transition suppression},
  author={Motzoi, Felix and Wilhelm, Frank K},
  journal={Phys. Rev. A},
  volume={88},
  number={6},
  pages={062318},
  year={2013},
  publisher={APS},
  doi={10.1103/PhysRevA.88.062318}
}

@article{nam2024,
  title={Reinforcement learning pulses for transmon qubit entangling gates},
  author={Nam Nguyen, Ho and Motzoi, Felix and Metcalf, Mekena and Birgitta Whaley, K and Bukov, Marin and Schmitt, Markus},
  journal={Machine Learning: Science and Technology},
  volume={5},
  number={2},
  pages={025066},
  year={2024},
  doi={10.1088/2632-2153/ad4f4d}
}

@article{oreg1984,
  title={Adiabatic following in multilevel systems},
  author={Oreg, J and Hioe, FT and Eberly, JH},
  journal={Phys. Rev. A},
  volume={29},
  number={2},
  pages={690},
  year={1984},
  publisher={APS},
  doi={10.1103/PhysRevA.29.690}
}

@article{preti2022,
  title={Continuous quantum gate sets and pulse-class meta-optimization},
  author={Preti, Francesco and Calarco, Tommaso and Motzoi, Felix},
  journal={PRX Quantum},
  volume={3},
  number={4},
  pages={040311},
  year={2022},
  doi={10.1103/PRXQuantum.3.040311}
}

@article{Rach2015,
  title={Dressing the chopped-random-basis optimization: A bandwidth-limited access to the trap-free landscape},
  author={Rach, N. and M{\"u}ller, M. M. and Calarco, T. and Montangero, S.},
  journal={Phys. Rev. A},
  volume={92},
  number={6},
  pages={062343},
  year={2015},
  doi={10.1103/PhysRevA.92.062343}
}

@article{sauvage2022,
  title={Optimal control of families of quantum gates},
  author={Sauvage, Fr{\'e}d{\'e}ric and Mintert, Florian},
  journal={Phys. Rev. Lett.},
  volume={129},
  number={5},
  pages={050507},
  year={2022},
  doi={10.1103/PhysRevLett.129.050507}
}

@article{singh2023,
  title={Compensating for nonlinear distortions in controlled quantum systems},
  author={Singh, Juhi and Zeier, Robert and Calarco, Tommaso and Motzoi, Felix},
  journal={Phys. Rev. Applied},
  volume={19},
  number={6},
  pages={064067},
  year={2023},
  doi={10.1103/PhysRevApplied.19.064067}
}

@article{vitanov2017,
  title={Stimulated Raman adiabatic passage in physics, chemistry, and beyond},
  author={Vitanov, Nikolay V and Rangelov, Andon A and Shore, Bruce W and Bergmann, Klaas},
  journal={Reviews of Modern Physics},
  volume={89},
  number={1},
  pages={015006},
  year={2017},
  doi={RevModPhys.89.015006}
}

@article{Watts2015,
  title={Optimizing for an arbitrary perfect entangler. I. Functionals},
  author={Watts, Paul and Vala, Jiri and M{\"u}ller, Matthias M. and Calarco, Tommaso and Whaley, K. Birgitta and Reich, Daniel M. and Goerz, Michael H. and Koch, Christiane P.},
  journal={Phys. Rev. A},
  volume={91},
  number={6},
  pages={062306},
  year={2015},
  doi={10.1103/PhysRevA.91.062306}
}

@article{Wineland1998TrappedIons,
  author  = {Wineland, D. J. and Monroe, C. and Itano, W. M.
             and Leibfried, D. and King, B. E.
             and Meekhof, D. M.},
  title   = {Experimental Issues in Coherent Quantum-State
             Manipulation of Trapped Atomic Ions},
  journal = {Journal of Research of the National Institute
             of Standards and Technology},
  volume  = {103},
  number  = {3},
  pages   = {259--328},
  year    = {1998},
  doi     = {10.6028/jres.103.019}
}

@article{Zhang2003a,
  title={Geometric theory of nonlocal two-qubit operations},
  author={Zhang, Jun and Vala, Jiri and Sastry, Shankar and Whaley, K. Birgitta},
  journal={Phys. Rev. A},
  volume={67},
  number={4},
  pages={042313},
  year={2003},
  publisher={APS},
 doi={10.1103/PhysRevA.67.042313}
}

@article{Zhang2003b,
  title={Exact Two-Qubit Universal Quantum Circuit},
  author={Zhang, Jun and Vala, Jiri and Sastry, Shankar and Whaley, K. Birgitta},
  journal={Phys. Rev. Lett.},
  volume={91},
  number={2},
  pages={027903},
  year={2003},
  publisher={APS},
 doi={10.1103/PhysRevLett.91.027903}
}

@article{Zhang2004a,
  title={Optimal quantum circuit synthesis from controlled-unitary gates},
  author={Zhang, Jun and Vala, Jiri and Sastry, Shankar and Whaley, K. Birgitta},
  journal={Phys. Rev. A},
  volume={69},
  number={4},
  pages={042309},
  year={2004},
  publisher={APS},
 doi={10.1103/PhysRevA.69.042309}
}

@article{unanyan1997,
  title={Laser-induced adiabatic atomic reorientation with control of diabatic losses},
  author={Unanyan, RG and Yatsenko, LP and Bergmann, K and Shore, BW},
  journal={Optics communications},
  volume={139},
  number={1-3},
  pages={48--54},
  year={1997},
  publisher={Elsevier},
  doi={10.1016/S0030-4018(97)00099-0}
}

@article{rigetti2010,
  title={Fully microwave-tunable universal gates in superconducting qubits with linear couplings and fixed transition frequencies},
  author={Rigetti, Chad and Devoret, Michel},
  journal={Phys. Rev. B—Condensed Matter and Materials Physics},
  volume={81},
  number={13},
  pages={134507},
  year={2010},
  publisher={APS},
  doi={10.1103/PhysRevB.81.134507}
}

@article{chow2012,
  title={Universal Quantum Gate Set Approaching Fault-Tolerant Thresholds with Superconducting Qubits},
  author={Chow, Jerry M and Gambetta, Jay M and Corcoles, Antonio D and Merkel, Seth T and Smolin, John A and Rigetti, Chad and Poletto, Stefano and Keefe, George A and Rothwell, Mary B and Rozen, John R and others},
  journal={Phys. Rev. Lett.},
  volume={109},
  number={6},
  pages={060501},
  year={2012},
  publisher={APS},
  doi={10.1103/PhysRevLett.109.060501}
}

@article{li2024,
  title={Experimental error suppression in Cross-Resonance gates via multi-derivative pulse shaping},
  author={Li, Boxi and Calarco, Tommaso and Motzoi, Felix},
  journal={npj Quantum Information},
  volume={10},
  number={1},
  pages={66},
  year={2024},
  publisher={Nature Publishing Group UK London},
  doi={10.1038/s41534-024-00863-4}
}

@article{caldwell2018,
  title={Parametrically activated entangling gates using transmon qubits},
  author={Caldwell, SA and Didier, N and Ryan, CA and Sete, EA and Hudson, A and Karalekas, P and Manenti, R and da Silva, MP and Sinclair, R and Acala, E ea and others},
  journal={Phys. Rev. Applied},
  volume={10},
  number={3},
  pages={034050},
  year={2018},
  publisher={APS},
  doi={10.1103/PhysRevApplied.10.034050}
}

@article{jaksch2000,
  title={Fast quantum gates for neutral atoms},
  author={Jaksch, Dieter and Cirac, Juan Ignacio and Zoller, Peter and Rolston, Steve L and C{\^o}t{\'e}, Robin and Lukin, Mikhail D},
  journal={Phys. Rev. Lett.},
  volume={85},
  number={10},
  pages={2208},
  year={2000},
  publisher={APS},
  doi={10.1103/PhysRevLett.85.2208}
}

@article{Theis2016,
  title = {High-fidelity Rydberg-blockade entangling gate using shaped, analytic pulses},
  author = {Theis, L. S. and Motzoi, F. and Wilhelm, F. K. and Saffman, M.},
  journal = {Phys. Rev. A},
  volume = {94},
  issue = {3},
  pages = {032306},
  numpages = {6},
  year = {2016},
  month = {Sep},
  publisher = {American Physical Society},
  doi = {10.1103/PhysRevA.94.032306}}

@article{jau2016,
  title={Entangling atomic spins with a Rydberg-dressed spin-flip blockade},
  author={Jau, Y-Y and Hankin, AM and Keating, Tyler and Deutsch, Ivan H and Biedermann, GW},
  journal={Nature Physics},
  volume={12},
  number={1},
  pages={71--74},
  year={2016},
  publisher={Nature Publishing Group UK London},
  doi={10.1038/nphys3487}
}

@article{renger2026,
  title={Superconducting qubit-resonator quantum processor with effective all-to-all connectivity},
  author={Renger, Michael and Verjauw, Jeroen and Wurz, Nicola and Hosseinkhani, Amin and Ockeloen-Korppi, Caspar and Liu, Wei and Rath, Aniket and Thapa, Manish J and Vigneau, Florian and Wybo, Elisabeth and others},
  journal={Phys. Rev. Research},
  volume={8},
  number={1},
  pages={013148},
  year={2026},
  publisher={APS},
  doi={10.1103/8x68-cx1w}
}

@article{blais2007,
  title={Quantum-information processing with circuit quantum electrodynamics},
  author={Blais, Alexandre and Gambetta, Jay and Wallraff, Andreas and Schuster, David I and Girvin, Steven M and Devoret, Michel H and Schoelkopf, Robert J},
  journal={Phys. Rev. A},
  volume={75},
  number={3},
  pages={032329},
  year={2007},
  publisher={APS},
  doi={10.1103/PhysRevA.75.032329}
}

@article{burkard1999,
  title={Coupled quantum dots as quantum gates},
  author={Burkard, Guido and Loss, Daniel and DiVincenzo, David P},
  journal={Phys. Rev. B},
  volume={59},
  number={3},
  pages={2070},
  year={1999},
  publisher={APS},
  doi={10.1103/PhysRevB.59.2070}
}

@article{childress2006,
  author = {Childress, L. and Gurudev Dutt, M. V. and Taylor, J. M. and Zibrov, A. S. and Jelezko, F. and Wrachtrup, J. and Hemmer, P. R. and Lukin, M. D.},
  title = {Coherent Dynamics of Coupled Electron and Nuclear Spin Qubits in Diamond},
  journal = {Science},
  volume = {314},
  number = {5797},
  pages = {281--285},
  year = {2006},
  doi = {10.1126/science.1131871}
}

@article{dicarlo2009,
  author = {DiCarlo, L. and Chow, J. M. and Gambetta, J. M. and Bishop, L. S. and Johnson, B. R. and Schuster, D. I. and Majer, J. and Blais, A. and Frunzio, L. and Girvin, S. M. and Schoelkopf, R. J.},
  title = {Demonstration of two-qubit algorithms with a superconducting quantum processor},
  journal = {Nature},
  volume = {460},
  number = {7252},
  pages = {240--244},
  year = {2009},
  doi = {10.1038/nature08121}
}

@article{strauch2003,
  title={Quantum logic gates for coupled superconducting phase qubits},
  author={Strauch, Frederick W and Johnson, Philip R and Dragt, Alex J and Lobb, CJ and Anderson, JR and Wellstood, FC},
  journal={Phys. Rev. Lett.},
  volume={91},
  number={16},
  pages={167005},
  year={2003},
  publisher={APS},
  doi={10.1103/PhysRevLett.91.167005}
}

@article{petta2005,
  author = {Petta, J. R. and Johnson, A. C. and Taylor, J. M. and Laird, E. A. and Yacoby, A. and Lukin, M. D. and Marcus, C. M. and Hanson, M. P. and Gossard, A. C.},
  title = {Coherent Manipulation of Coupled Electron Spins in Semiconductor Quantum Dots},
  journal = {Science},
  volume = {309},
  number = {5744},
  pages = {2180--2184},
  year = {2005},
  doi = {10.1126/science.1116955}
}

@article{jaksch1999,
  author = {Jaksch, D. and Briegel, H.-J. and Cirac, J. I. and Gardiner, C. W. and Zoller, P.},
  title = {Entanglement of Atoms via Cold Controlled Collisions},
  journal = {Phys. Rev. Lett.},
  volume = {82},
  number = {9},
  pages = {1975--1978},
  year = {1999},
  doi = {10.1103/PhysRevLett.82.1975}
}

@article{calarco2000,
  author = {Calarco, T. and Hinds, E. A. and Jaksch, D. and Schmiedmayer, J. and Cirac, J. I. and Zoller, P.},
  title = {Quantum gates with neutral atoms: Controlling collisional interactions in time-dependent traps},
  journal = {Phys. Rev. A},
  volume = {61},
  number = {2},
  pages = {022304},
  year = {2000},
  doi = {10.1103/PhysRevA.61.022304}
}

@article{gorini1976,
  author = {Gorini, Vittorio and Kossakowski, Andrzej and Sudarshan, E. C. G.},
  title = {Completely Positive Dynamical Semigroups of N-Level Systems},
  journal = {Journal of Mathematical Physics},
  volume = {17},
  number = {5},
  pages = {821--825},
  year = {1976},
  doi = {10.1063/1.522979}
}

@article{lindblad1976,
  author = {Lindblad, Goran},
  title = {On the Generators of Quantum Dynamical Semigroups},
  journal = {Communications in Mathematical Physics},
  volume = {48},
  number = {2},
  pages = {119--130},
  year = {1976},
  doi = {10.1007/BF01608499}
}

@article{rabi1937,
  author = {Rabi, I. I.},
  title = {Space Quantization in a Gyrating Magnetic Field},
  journal = {Physical Review},
  volume = {51},
  number = {8},
  pages = {652--654},
  year = {1937},
  doi = {10.1103/PhysRev.51.652}
}

@article{bloch1940,
  author = {Bloch, Felix and Siegert, Arnold},
  title = {Magnetic Resonance for Nonrotating Fields},
  journal = {Physical Review},
  volume = {57},
  number = {6},
  pages = {522--527},
  year = {1940},
  doi = {10.1103/PhysRev.57.522}
}

@article{autler1955,
  author = {Autler, S. H. and Townes, C. H.},
  title = {Stark Effect in Rapidly Varying Fields},
  journal = {Physical Review},
  volume = {100},
  number = {2},
  pages = {703--722},
  year = {1955},
  doi = {10.1103/PhysRev.100.703}
}

@article{magnus1954,
  author = {Magnus, Wilhelm},
  title = {On the Exponential Solution of Differential Equations for a Linear Operator},
  journal = {Communications on Pure and Applied Mathematics},
  volume = {7},
  number = {4},
  pages = {649--673},
  year = {1954},
  doi = {10.1002/cpa.3160070404}
}

@article{zener1932,
  author = {Zener, Clarence},
  title = {Non-Adiabatic Crossing of Energy Levels},
  journal = {Proceedings of the Royal Society of London. Series A},
  volume = {137},
  number = {833},
  pages = {696--702},
  year = {1932},
  doi = {10.1098/rspa.1932.0165}
}

@article{kato1950,
  author = {Kato, Tosio},
  title = {On the Adiabatic Theorem of Quantum Mechanics},
  journal = {Journal of the Physical Society of Japan},
  volume = {5},
  number = {6},
  pages = {435--439},
  year = {1950},
  doi = {10.1143/JPSJ.5.435}
}

@article{muga2010,
  author = {Muga, J. G. and Chen, X. and Ibanez, S. and Lizuain, I. and Ruschhaupt, A.},
  title = {Transitionless Quantum Drivings for the Harmonic Oscillator},
  journal = {Journal of Physics B: Atomic, Molecular and Optical Physics},
  volume = {43},
  pages = {085509},
  year = {2010},
  doi = {10.1088/0953-4075/43/8/085509}
}

@article{watkins1992,
  author = {Watkins, Christopher J. C. H. and Dayan, Peter},
  title = {Q-Learning},
  journal = {Machine Learning},
  volume = {8},
  number = {3},
  pages = {279--292},
  year = {1992},
  doi = {10.1007/BF00992698}
}

@misc{schulman2017,
  author = {Schulman, John and Wolski, Filip and Dhariwal, Prafulla and Radford, Alec and Klimov, Oleg},
  title = {Proximal Policy Optimization Algorithms},
  year = {2017},
  eprint = {1707.06347},
  archivePrefix = {arXiv},
  doi = {10.48550/arXiv.1707.06347}
}

@misc{lillicrap2015,
  author = {Lillicrap, Timothy P. and Hunt, Jonathan J. and Pritzel, Alexander and Heess, Nicolas and Erez, Tom and Tassa, Yuval and Silver, David and Wierstra, Daan},
  title = {Continuous Control with Deep Reinforcement Learning},
  year = {2015},
  eprint = {1509.02971},
  archivePrefix = {arXiv},
  doi = {10.48550/arXiv.1509.02971}
}

@article{silver2018,
  author = {Silver, David and Hubert, Thomas and Schrittwieser, Julian and Antonoglou, Ioannis and Lai, Matthew and Guez, Arthur and Lanctot, Marc and Sifre, Laurent and Kumaran, Dharshan and Graepel, Thore and Lillicrap, Timothy and Simonyan, Karen and Hassabis, Demis},
  title = {A General Reinforcement Learning Algorithm That Masters Chess, Shogi, and Go Through Self-Play},
  journal = {Science},
  volume = {362},
  number = {6419},
  pages = {1140--1144},
  year = {2018},
  doi = {10.1126/science.aar6404}
}

@article{jurdjevic1972,
  author = {Jurdjevic, Velimir and Sussmann, Hector J.},
  title = {Control Systems on Lie Groups},
  journal = {Journal of Differential Equations},
  volume = {12},
  number = {2},
  pages = {313--329},
  year = {1972},
  doi = {10.1016/0022-0396(72)90035-6}
}

@article{altafini2002,
  author = {Altafini, Claudio},
  title = {Controllability of Quantum Mechanical Systems by Root Space Decomposition of su(N)},
  journal = {Journal of Mathematical Physics},
  volume = {43},
  number = {5},
  pages = {2051--2062},
  year = {2002},
  doi = {10.1063/1.1467611}
}

@article{barenco1995,
  author = {Barenco, Adriano and Bennett, Charles H. and Cleve, Richard and DiVincenzo, David P. and Margolus, Norman and Shor, Peter and Sleator, Tycho and Smolin, John A. and Weinfurter, Harald},
  title = {Elementary Gates for Quantum Computation},
  journal = {Physical Review A},
  volume = {52},
  number = {5},
  pages = {3457--3467},
  year = {1995},
  doi = {10.1103/PhysRevA.52.3457}
}

@article{pezze2018,
  author = {Pezz{\`e}, Luca and Smerzi, Augusto and Oberthaler, Markus K. and Schmied, Roman and Treutlein, Philipp},
  title = {Quantum Metrology with Nonclassical States of Atomic Ensembles},
  journal = {Reviews of Modern Physics},
  volume = {90},
  pages = {035005},
  year = {2018},
  doi = {10.1103/RevModPhys.90.035005}
}

@article{norambuena2024,
  author  = {Norambuena, Ariel and Mattheakis, Marios and
             Gonz{\'a}lez, Francisco J. and Coto, Ra{\'u}l},
  title   = {Physics-Informed Neural Networks for Quantum Control},
  journal = {Physical Review Letters},
  volume  = {132},
  number  = {1},
  pages   = {010801},
  year    = {2024},
  doi     = {10.1103/PhysRevLett.132.010801}
}

@article{khalid2023,
  author  = {Khalid, Irtaza and Weidner, Carrie A. and
             Jonckheere, Edmond A. and Schirmer, Sophie G. and
             Langbein, Frank},
  title   = {Sample-efficient model-based reinforcement learning
             for quantum control},
  journal = {Physical Review Research},
  volume  = {5},
  number  = {4},
  pages   = {043002},
  year    = {2023},
  doi     = {10.1103/PhysRevResearch.5.043002}
}

@book{sutton2018,
  author    = {Sutton, Richard S. and Barto, Andrew G.},
  title     = {Reinforcement Learning: An Introduction},
  edition   = {2},
  publisher = {MIT Press},
  address   = {Cambridge, MA},
  year      = {2018},
  isbn      = {9780262039246}
}
	
\end{document}